\documentclass[12pt,preprintnumbers,nofootinbib,notitlepage,aps,pra,showpacs]{revtex4-1}
\usepackage{amsmath,amssymb}
\usepackage{slashed,verbatim,graphicx}
\usepackage[colorlinks=true,citecolor=blue,linkcolor=blue]{hyperref}
\usepackage{color}
\usepackage[normalem]{ulem}
\usepackage[english]{babel}

\newcommand{\rmv}[1]{}
\newcommand{\keep}[1]{#1}

\newcommand{\vac}{|0\rangle}

\newcommand\ket[1]{| #1\rangle}
\newcommand\bra[1]{\langle #1|}
\newcommand\braket[2]{\langle #1|#2\rangle}

\def\Tr{\textrm{Tr}}
\newcommand{\CH}{\mathcal{H}}

\newcommand{\beq}{\begin{equation}}
\newcommand{\beqs}{\begin{equation*}}
\newcommand{\eeq}{\end{equation}}
\newcommand{\eeqs}{\end{equation*}}

\begin{document}
\setlength{\unitlength}{1mm}


\title{Entanglement entropy converges to classical entropy around periodic orbits}


\author{Curtis T. Asplund}
\affiliation {Department of Physics, Columbia University, 538 West 120th Street, New York, New York 10027}
\email{ca2621@columbia.edu}

\author{David Berenstein}
\affiliation{Department of Applied Mathematics and Theoretical Physics, 
University of Cambridge, Wilberforce Road, Cambridge CB3 0WA, United Kingdom}
\altaffiliation{Permanent address: Department of Physics, University of California Santa Barbara, Santa Barbara, California 93106}
\email{dberens@physics.ucsb.edu}

\preprint{DAMTP-2015-15}
\pacs{03.67.Bg, 05.45.Mt}

\begin{abstract}
We consider oscillators evolving subject to a periodic driving force that dynamically entangles them, and argue that this gives the linearized evolution around periodic orbits in a general chaotic Hamiltonian dynamical system. We show that the entanglement entropy, after tracing over half of the oscillators, generically asymptotes to linear growth at a rate given by the sum of the positive Lyapunov exponents of the system. These exponents give a classical entropy growth rate, in the sense of Kolmogorov, Sinai and Pesin. We also calculate the dependence of this entropy on linear mixtures of the oscillator Hilbert space factors, to investigate the dependence of the entanglement entropy on the choice of coarse-graining. We find that for almost all choices the asymptotic growth rate is the same.
\end{abstract}

\maketitle


\section{Introduction}
\label{S:Introduction}

\rmv{Various results in quantum information theory} \keep{The results of \cite{Page:1993df} state that, roughly speaking,}\rmv{suggest that} if we have a quantum system 
with a finite-dimensional Hilbert space of states that we factorize into a product of Hilbert spaces,
\begin{equation}
{ \CH}= {\CH}_A\otimes {\CH}_B \ ,
\end{equation}
then the typical pure state in ${\cal H}$ has very close to the maximal amount of entanglement allowed between ${\cal H}_A$ and ${\cal H}_B$,  
and this is in turn maximized if  $\dim( \CH_A)=\dim (\CH_B)$. 
We will call such factorization of $\CH$ into ``observable" and ``non-observable" physics a coarse-graining of the system.
This suggests that if we evolve a system randomly from an initial configuration with zero entanglement entropy, 
then\rmv{time evolution} \keep{it} will eventually forget essentially all of the information of the initial state if we only measure 
observables sensitive to $\CH_A$. 
\rmv{Moreover, it has been argued that the entropy growth is linear in time until 
it reaches saturation.}
There are many studies of this kind of process in specific systems.
The rate at which the entanglement grows towards saturation depends, in 
general, on the details of these systems, 
although some general bounds exist \cite{2007PhRvA..76e2319B, 2013PhRvL.111q0501V,Avery:2014dba, Bousso:2014uxa}.
Linear growth in time appears in many systems, for example, studies of decoherence and quantum chaos  
\cite{Zurek:1994wd, Zurek:1995jd, PhysRevLett.83.4526, PhysRevE.60.1542, 2000PhRvL..85.3373M, 
2002PhRvE..66d5201T, *2003PhRvE..67f6201F, *2003JPSJ...72S.111F, 2003PhRvA..68c2104B, 2003JPhA...36.2463Z, 
2004JPhA...37.5157A, 2004PhRvL..92o0403J, *PhysRevLett.93.219903, 2006PhRvL..97s4103P, 2009AdPhy..58...67J, 2008JPhA...41k5303F, 
2014OptCo.331..148S, 2010PhRvA..82e2335R, 2011PhRvE..83d6214B},
and quenches of extended systems \cite{Calabrese:2005in, Calabrese:2007rg, Hubeny:2007xt, AbajoArrastia:2010yt}.

When trying to\rmv{use this quantum information result} \keep{apply these results} in the context of black hole physics, 
we are usually confronted with two problems. First of all, the Hilbert space ${\CH}$ is big. 
In the gauge/gravity duality \cite{Maldacena:1997re} the dynamics takes place in an infinite-dimensional Hilbert space: 
it is the Hilbert space of a relativistic quantum field theory on the conformal boundary.

A very naive application of the results of \cite{Page:1993df} would suggest that typical states have infinite entropy 
when splitting $\CH$ in two pieces of the same size, since both are infinite dimensional. 
However, the notion of splitting along a random factorization has no meaning, 
because once we have factored into infinite dimensional pieces, we can
factorize the pieces again: there is no natural notion of splitting in half. 
Thus, the question of the entanglement entropy for a typical state is 
ill-defined without additional structure on the Hilbert space. 

\rmv{Given such a structure, one would obtain}
\keep{An example of such a structure is} two operators algebras, one for $\CH_A$ the other one for $\CH_B$.
It is natural to do the splitting with respect to a choice of algebras with\rmv{some}  reasonable 
properties determined by\rmv{some} feature\keep{s} of the dynamics.  
 Once that splitting is done,\rmv{what we can do} instead of computing the entanglement entropy of the typical state, 
 we can compute the rate of growth\rmv{in} \keep{of} the entanglement entropy as governed by the dynamics and ask how this growth is affected by our 
 choices of coarse-graining and dynamics. It is here that we need a model dynamics that is both tractable and generic. \keep{We have in mind two simple harmonic oscillators with two ladder operator 
algebras, and we will show in what sense a system like this can be considered generic.}

The second problem we find generically is that there is no  obvious canonical splitting into 
two factors $\CH_A$ and $\CH_B$, 
so one might expect that entanglement entropy\rmv{estimates} based on some such splitting might depend substantially on the 
coarse-graining\rmv{choices that are made}. 
If we also define the scrambling\rmv{times} rate as the slope of the entropy growth, 
one might worry that there is no objective way to quantify \rmv{scrambling}\keep{it}.
This would make it very hard to understand in what sense black holes are fast scramblers 
\cite{Hayden:2007cs,Sekino:2008he}, when we think of the evolution in terms of a dual quantum field theory. In this sense,
it is natural to ask if there is a universal result where the details of the factorization 
do not matter too much. 
Our main motivation  is to eventually formulate the fast scrambling conjecture on a rigorous footing, 
but to do so, we need to be able to apply the methods that could characterize 
scrambling to fairly generic dynamical systems to which we associate an infinite-dimensional Hilbert space.

 The purpose of this paper is to analyze the scrambling rate, i.e., 
 the entropy growth rate, in a toy model of chaotic dynamics 
that iterates a relatively simple unitary evolution operator on an infinite-dimensional Hilbert space $\CH$. 
\keep{The Hilbert space} will be further decomposed into a product of two infinite-dimensional Hilbert spaces. 
This is done for a closed system, and we will study the dependence on the choice of coarse-graining and initial state. 
The idea is to study the quantum counterparts of closed Hamiltonian 
chaotic dynamical systems with finitely many degrees of freedom, similarly to previous 
studies of closed systems in the quantum chaos and decoherence literature \cite{PhysRevE.60.1542, 2003PhRvA..68c2104B, 2008JPhA...41k5303F}.

So long as these classical systems have bounded trajectories (for example if the regions with bounded energy have finite volume), 
then \rmv{it is}\keep{they are} expected \rmv{that such systems will}\keep{to} have a dense set of periodic trajectories. 
\rmv{Our simplification is to}\keep{We} assume that these periodic trajectories encode all the important information of the dynamical system, 
and that any classical initial condition \rmv{can be understood as being}\keep{is} sufficiently close to such a periodic orbit, 
in line with periodic-orbit theory in quantum chaos \cite{Gutzwiller:1971fy, Gutzwiller1991, Cvitanovic2012}. 
The evolution of the classical system for such an initial condition 
can \keep{then} be understood, for some time, by the linearized perturbations around the corresponding periodic orbit. 
We can ask how these 
perturbations grow in time and\rmv{argue that we can} estimate the Lyapunov exponents of the full system 
from such an analysis. 
Given such Lyapunov exponents, one then has the
classical entropy growth \keep{rate}, 
essentially the Kolmogorov-Sinai (KS) entropy.\footnote{See \cite{Cvitanovic2012,Zaslavsky2005,Zaslavsky2007} for general discussions on chaotic Hamiltonian dynamics 
and more precise definitions. 
The technical definition of the KS entropy requires a finite measure space, while we are working with a phase space
and Liouville measure that are only $\sigma$-finite, but our results do not depend on this issue.} 
 
 The main point of this paper is that the entropy growth rate is the same in the classical and quantum systems 
and \keep{that it is} essentially independent of the coarse-graining and  
initial state (\keep{assuming} that the latter is a sufficiently good approximation to a classical state).
Our results\rmv{essentially match} \keep{are consistent with} the seminal work on open systems by Zurek and Paz \cite{Zurek:1994wd}, 
where a preferred factorization into a degree of freedom and an environment is given, 
but note that we explicitly find that the\rmv{rates of entropy production} 
\keep{entanglement entropy growth rate} 
grows as the sum of the positive Lyapunov exponents, not just \keep{as} the largest one. 
The original work of Zurek and Paz \cite{Zurek:1994wd} had only one such positive exponent, 
and similarly with \cite{2003PhRvA..68c2104B, 2008JPhA...41k5303F},
though other studies have also found growth rates equal to the full sum \cite{Zurek:1995jd, PhysRevE.60.1542}.
We are also consistent with other earlier studies of entanglement entropy growth, 
but distinguished from them in that our system is not coupled to an external environment 
\cite{Zurek:1994wd, Zurek:1995jd, PhysRevLett.83.4526, 2000PhRvL..85.3373M, 2004JPhA...37.5157A, 2011PhRvE..83d6214B, 2014OptCo.331..148S},
our results are not perturbative \cite{2003JPhA...36.2463Z} or numerical \cite{PhysRevLett.80.5524, 2012PhRvE..85c6208C}
and we do not work with a finite-dimensional Hilbert space 
\cite{PhysRevE.60.1542, 2002PhRvE..66d5201T, *2003PhRvE..67f6201F, *2003JPSJ...72S.111F} or 
invoke a random matrix or semi-classical approximation 
\cite{2005PhRvA..71d2321A, 2004PhRvL..92o0403J, *PhysRevLett.93.219903, 2006PhRvL..97s4103P, 2009AdPhy..58...67J, 2010PhRvA..82e2335R}. 

We also go further in that we explicitly show that the choice of coarse graining in this closed system 
\keep{almost always} matters very little\rmv{except for sets of measure zero on the choices that we allow ourselves to make}.  
We consider all factorizations of the Hilbert space in which each factor corresponds to half the 
degrees of freedom.\footnote{More precisely, over a factor spanned by the states generated by the quantization of a single particle coordinate $\hat{q}$ and 
its canonical conjugate $\hat{p}$, which we combine into ladder operators $\hat{q} \pm i\hat{p}$.}
This is the closest analog we can imagine in this class of systems to saying that $\CH_A$ and $\CH_B$ are the same size 
(since both are infinite dimensional), while ensuring that the observables in $\CH_A$ commute with the observables in $\CH_B$. 
We focus on the minimal system for which this splitting can be done, namely, 
\rmv{a collection of two pairs of phase space coordinates $x,p$}\keep{two degrees of freedom}, 
but\rmv{it is clear that} the methods\rmv{illustrated} \keep{used} here can be generalized to more 
degrees of freedom\rmv{in a relatively straightforward way}.


\section{A simple stroboscopic dynamics}\label{sec:strobo}

As described in the introduction, 
\rmv{our goal is to estimate}\keep{we are interested in} the entanglement entropy growth for a quantum system with an infinite-dimensional Hilbert space after acting repeatedly on it with a unitary operator on a preferred initial pure state. The quantum system is supposed to arise\rmv{after} \keep{from} 
quantization of a classical (chaotic) dynamical system\rmv{with an a-priori known dynamics}.
We consider an initial pure 
state\rmv{is required to be} as close to a classical initial condition as\rmv{is} possible given the uncertainty principle,
\keep{like a coherent state of a harmonic oscillator,} since 
this should allow us to clearly see the relationship between classical and 
quantum entropy growth.

\rmv{This calculation, as a general setup, seems too abstract.}
We are not able to solve this problem in general, so
we solve a highly simplified model as a first step.\rmv{that we can analyze more readily.}
\rmv{The basic idea is to consider that}In chaotic Hamiltonian dynamical systems with bounded 
phase space trajectories, 
e.g., those whose constant-energy surfaces are compact, 
there are many periodic solutions\rmv{to the equations of motion} (in highly chaotic 
situations they are dense in the phase space), 
and we will choose our initial condition to lay very close to one such periodic trajectory, 
with period $T$, but in a chaotic domain of phase space, i.e., away from invariant 
(Kolmogorov-Arnol'd-Moser) tori.
If we choose such an initial condition and wait a time $T$, we will end up\rmv{very} close to our initial condition and we 
have a canonical transformation that has a fixed point set: 
the periodic trajectory\rmv{themselves} itself. 
Other nearby periodic trajectories will have much longer periods in general, 
so we can ignore them if we only study the system at times comparable to $T$ or perhaps one order of magnitude larger.
This is all we need for most of our results to hold, but see the end of 
Sec.~\ref{sec:Ent1} for a more quantitative discussion.

The simplification of the dynamics occurs by 
considering an infinitesimal neighborhood about some point on this periodic orbit,
and assuming that this can be treated 
with a linearized, tangent-space approximation to the dynamics. 
This approximation should keep all the essential features of the dynamics in an expanded,
finite neighborhood, by the Hartman-Grobman linearization theorem.

Note that by translating the solution by a small amount of time again leads to an exactly 
periodic solution with period $T$, so there is no stretching in this direction. 
Geometrically, this is the direction along the periodic orbit that pierces the 
small neighborhood we are considering. 
This direction that experiences no growth leads to one vanishing Lyapunov exponent. 
There is a second direction that experiences no growth, which is related to conservation of energy. These two 
 form a conjugate pair.
We will consider only directions linearly independent of these directions, and 
so will generally have an even number of non-zero Lyapunov exponents.\

Thus, we are led to consider a simplified problem where we have a linear canonical transformation in phase space 
such that\rmv{the image of the origin is the origin} \keep{the origin is fixed}. 
Eventually we want to treat the system quantum mechanically, 
and we will form a quantization of this linear phase space in terms of operators 
$\hat q_s,\hat p_s$ that transform linearly into each other.
We do this by first choosing a set of Darboux coordinates, $q_s,p_s$, centered on the fixed point, 
that parametrize the tangent space of the fixed point
and simplify the symplectic form to $\Omega=\sum_s dq_s\wedge dp_s$. 
Then we can associate a copy of the \keep{ladder operator} algebra with each pair of canonical coordinates:
\beq
\label{eq:ladder}
a_s= (\hat{q}_s +i \hat{p}_s)/\sqrt{2}\quad \text{and}\quad a_s^\dagger = (\hat{q}_s-i \hat{p}_s)/\sqrt{2}\ .
\eeq

\keep{Evolution in our model is effected by a canonical linear transformation, which} 
can\rmv{therefore} be\rmv{understood as} \keep{represented by} a real matrix that preserves $\Omega$. 
In matrix notation, preserving $\Omega$ corresponds to the equation
\begin{equation}
M^T\Omega M = \Omega \ , \label{eq:pres}
\end{equation}
where we choose to represent $\Omega$ as an antisymmetric matrix and $M^T$ is the transpose of $M$.
We can then decompose the linearized phase space into 
subspaces that diagonalize $M$.

Assuming that we have an eigenvalue $\lambda$, since $M$ is real, we must have that $\bar\lambda$ is also an eigenvalue. Furthermore, preserving $\Omega$ can also be written as 
\begin{equation}
M^T= \Omega M^{-1} \Omega^{-1} \ ,
\end{equation}
which shows that $M^T$ is conjugate to $M^{-1}$. Since the transpose of a 
matrix and the matrix have the same set of eigenvalues, any eigenvalue of  $M^T$ corresponds to an eigenvalue of $M$. 
The eigenvalues of $M^T$ are the inverses of the eigenvalues of $M$. 
We thus find that to any eigenvalue, we can\rmv{in principle} associate a four-dimensional subspace of phase space,
the (possibly complexified) linear space spanned by the eigenvectors with eigenvalues $\lambda, \bar \lambda, \lambda^{-1}, \bar\lambda^{-1}$,
which are generically distinct. 
The eigenvalues cannot be zero because $\Omega$ is non-degenerate.

The action of $\Omega$ is closed on these linear subspaces. Consider two vectors $v_1, v_2$ that are eigenvectors of $M$. Then we have that
\begin{equation}
v_2^T \Omega M v_1= \lambda_1 v_2^T \Omega v_1= v_2^T (M^{-1})^T \Omega v_1= (M^{-1} v_2)^T \Omega v_1=
\lambda_2^{-1} v_2^T \Omega v_1 \ ,
\end{equation}
so if the matrix element of $v_2^T \Omega v_1$ is non-zero, we find that $\lambda_2= \lambda_1^{-1}$. 
That is, we find that the
\keep{linearized} phase space\rmv{under the action of $M$} can 
\keep{generically} be decomposed\rmv{in groups of variables made generically of} 
\keep{into four-dimensional} linear subspaces\rmv{of dimension 4}, 
such that the action of $M$ and $\Omega$ closes on these subspaces, 
so long as there are no degeneracies \keep{in the eigenvalues}.
\rmv{A degeneracy is the equality of two such possible eigenvalues.} 

There are three different cases of degeneracy to consider.
First, assume that $\lambda=\bar \lambda$. Then the eigenvalues are real, 
and the\rmv{phase space factor that has been identified} \keep{subspace} is only two dimensional. 
These will be called unstable directions, because upon iteration the distance from the origin will grow indefinitely.
The second possibility is that $\lambda=\bar \lambda^{-1}$, which also gives a two dimensional slice of phase space. 
Then the eigenvalues lie on the unit circle, and the direction is said to be stable.
Finally, assume  $\lambda_1= \lambda_1^{-1}$, in which case $\lambda=\pm 1$. 
These necessarily come in pairs of eigenvalues because $\Omega$ is antisymmetric and non-degenerate, 
so there must exist a vector $v_2$ to any eigenvector $v_1$ such that $v_2^T \Omega v_1\neq 0$.
By the argument above, if $v_2$ is an eigenvector of $M$ itself,  it has the same eigenvalue.

This third case lies at the intersection of stable and unstable subspaces. 
These will be called marginally unstable. It is in this case that more exotic things can happen
because the matrix $M$ might not be diagonalizable, though it can always be put in 
upper-triangular form. 
A simple example is a two-dimensional phase space where we get the transformation
\begin{equation}
M\simeq \begin{pmatrix} 1& a \\ 0&1\end{pmatrix}
\end{equation}
which has two identical eigenvalues, but the matrix does not have two eigenvectors.
Iterations of the transformation lead to a constant drift
\begin{equation}
q_n=  q_0 + a n p
\end{equation}
depending on the conjugate momentum.
This corresponds to sub-exponential growth of the distance between initially nearby trajectories. 
These blocks are of vanishing measure in the set of all possible 
linear canonical transformations, but they might appear as non-trivial limits.  
It is an example like this that corresponds to the pair of coordinates consisting of energy plus the direction of flow along the periodic orbit. The role of $p$ is played by the energy, while $q_0$ is the linearized coordinate along the flow.

This is also relevant for 
integrable systems with bounded orbits, which are characterized by having all directions stable or marginally unstable. 
For the rest of this paper we work in a generic case with no degeneracy.

%
%

\section{Quantum linear dynamics for two degrees of freedom with a conserved $\text{U}(1)$ charge.}

The entanglement entropy is calculated by first factorizing the infinite-dimensional Hilbert space,
generated by the above ladder operators (see Eq.~\eqref{eq:ladder}), into two infinite-dimensional factors. 
\keep{We are effectively treating the $\hat{q}_s$ and $\hat{p}_s$ operators as position and momentum operators 
of harmonic oscillators.}
We will trace over \rmv{the other half}\keep{one factor} to compute the entanglement entropy.\rmv{of the subsystem.}
The idea \rmv{will be}\keep{is} to show that the classical entropy rate 
(the sum of the positive Lyapunov exponents\rmv{or Kolmogorov-Sinai entropy})
\keep{is equal to} 
the growth rate of the entanglement entropy for 
late times, for almost all choices of coarse-graining (factorization).

To simplify matters further, we will require that\rmv{additionally} the phase space be only 
of dimension four.\rmv{(this is, it has four real dimensions).} That is, we consider the phase space  
to be described by two $q,p$ pairs of canonical coordinates, 
or in the quantum theory, by two harmonic oscillator algebras. 
As a further simplification, \keep{we will impose a $\text{U}(1)$ symmetry},
so that the linear action preserves a $\text{U}(1)$ charge, 
with the two raising operators having opposite charge.\footnote{This will 
effectively reduce the possible evolution operators to consider from all of 
$\text{Sp}(4,\mathbb{R})$ (this includes consistency with the adjoint operation) 
to the subgroup $\text{SL}(2,\mathbb{R})$, as we see in more detail 
below.}
This choice is partly from our desire to 
understand entropy production near black hole horizons \`{a} la Hawking \cite{Hawking:1974sw}, 
where there is a distinction between particles and antiparticles in terms of modes that fall into or 
escape from a black hole.\rmv{versus modes that exit from it.} 
The ones that fall into the black hole are considered unobservable and one needs to trace over them. 
In the end, the main advantage of the $\text{U}(1)$ symmetry is that it leads 
to \keep{simple but illustrative} computations.
\rmv{that can be very simple to perform and which
that are illustrative of the general dynamics.}

Furthermore, this requirement of the $\text{U}(1)$ symmetry is not as constraining as it might seem. 
If we assign a charge $1$ to the coordinate whose eigenvalues
under the evolution matrix $M$ is $\lambda$,  we can also assign charge $+1$ to the coordinate associated to the eigenvalue $\bar{\lambda}^{-1}$, and charges $-1$ to the other two coordinates, associated with the eigenvalues 
$\bar{\lambda}$ and $\lambda^{-1}$. 
This charge is preserved by the canonical transformation. 
Thus, a conserved charge can be defined
whenever $\lambda$ is a generic complex number.

Let us now write the most general form of such a linear dynamics in the quantum theory, 
using two ladder operator algebras, $a, a^\dagger, b, b^\dagger$, with both $a^\dagger $ and $b$ of charge $1$, 
and working in the Heisenberg picture of quantum dynamics.
The most general 
\rmv{mixing under such a} canonical transformation will take $a^\dagger$ to a linear 
combination of $a^\dagger$ and $b$, and similarly, $b$ will be taken to a linear combination of $b$ and $a^\dagger$. For example,
\begin{equation}
a^\dagger = A a^\dagger_{\text{New}}+B b_{\text{New}} \ .
\end{equation} 
The most general such transformation can be parametrized as follows (see Appendix \ref{ap:bog} for the derivation)
\begin{equation}
\begin{pmatrix}
a^\dagger\\
b
\end{pmatrix}= \begin{pmatrix}
A&B\\
C&D
\end{pmatrix} \begin{pmatrix}
a^\dagger_{\text{New}}\\
b_{\text{New}}
\end{pmatrix} =\exp(i\beta) \begin{pmatrix} e^{i \theta} \cosh\rho & e^{i\phi}\sinh\rho \\
 e^{-i\phi}\sinh\rho & e^{-i \theta}\cosh \rho \end{pmatrix}\begin{pmatrix}
a^\dagger_{\text{New}}\\
b_{\text{New}}
\end{pmatrix} \label{eq:param}
\end{equation}
The transformation of $a, b^\dagger$ follows from taking hermitian conjugates. This is\rmv{called} a Bogolubov transformation.
\rmv{when written in terms of the oscillator algebras.}

This shows\rmv{clearly} that the associated group is a $\text{U}(1) \times \text{SL}(2,\mathbb{R})$, 
where the $\text{U}(1) $ corresponds to the phase $\exp(i\beta)$. Indeed, if we forget about $\beta$, the matrix in 
Eq.~\eqref{eq:param} has determinant equal to one, and trace given by $t=2\cosh \rho \cos\theta$. 
The eigenvalues of that matrix are solutions of 
\begin{equation}
\lambda^2 -t \lambda +1=0 \ . \label{eq:eivals}
\end{equation}
There are two cases to consider: $|t|>2$ and $|t|\leq 2$. In the first case the eigenvalues are real and inverses of each other, so we parametrize them as $\pm\exp(\hat \rho)$ and $\pm\exp(-\hat \rho)$, whereas in the other case the eigenvalues are on the unit circle and are $\exp(\pm i \hat\theta)$ . Therefore we write
$t= \pm 2 \cosh \hat \rho$ or $t= 2 \cos \hat \theta$ for both cases. 

For the full matrix we need to multiply both  eigenvalues by $\exp(i\beta)$, thus we can make sure that $t>0$ by changing the sign of $\exp(i\beta)$ if necessary. When the eigenvalues are on the unit circle, the system is said to be stable. We are mostly interested in understanding the unstable case, but we will work in the general setup anyhow.

The next thing we need to do is to choose an initial state. This will be the vacuum of the $a,b$ oscillators, characterized uniquely by 
\begin{equation}
a\vac _0= b\vac _0 =0 \ .
\end{equation}
We want to know what the new state will be after application of the canonical transformation. The answer follows from expressing
\begin{equation}
b\vac_0 = (D b_{\text{New}} + Ca_{\text{New}}^\dagger) \vac_0=0 \ .
\end{equation}
The state can be interpreted as an ``eigenvector" of the lowering operator $b_{\text{New}}$, with an operator valued eigenvalue proportional to $a_{\text{New}}^\dagger$, so it can be written in a similar form  as we do with coherent states, namely
\begin{equation}
\vac_0 \propto \exp ( -CD^{-1} a^\dagger b^\dagger)\vac_1 \label{eq:squeeze}
\end{equation}
and in general
\begin{equation}
\vac_0\propto \exp(-C_n D_n^{-1} a^\dagger b^\dagger)\vac_n \label{eq:iter}
\end{equation}
Notice that this expression is independent of $\beta$, so $\beta$\rmv{essentially} plays no significant role in the rest of the paper.

Our notation is as follows. The vacua $\vac_n$ will denote the vacuum of the oscillator algebra $a,a^\dagger, b, b^\dagger$ after $n$ iterations of the unitary evolution characterized by the matrix $M$. The initial state $\vac_0$ is  a squeezed state in terms of the new oscillator algebra after $n$ iterations. The preserved charge is 
\begin{equation}
Q=a^\dagger a-b^\dagger b\label{eq:Q}
\end{equation}
we will say that $a^\dagger$ creates particles and $b^\dagger$ creates antiparticles just as is done in conventional quantum field theory.

To choose different initial states, we do another Bogolubov transformation to a different set of raising and lowering operators at time zero, such that the Bogolubov transformation preserves the charge. This will change the matrix $M$ by conjugation, but the eigenvalues of $M$ will stay the same. The new state will be the vacuum of the new set of oscillators, which can be interpreted as an initial state which is a squeezed state in terms of the old algebra. All of the
states we have chosen as initial states are minimum uncertainty states centered at the fixed point.

Notice that if we fix $t=2 \cosh \hat \rho$, there is a particularly simple matrix with the right characteristic polynomial. This matrix is
\begin{equation}
M=\begin{pmatrix}
A&B\\
C&D
\end{pmatrix}= \begin{pmatrix}\cosh \hat \rho& \sinh\hat \rho\\
\sinh\hat \rho& \cosh\hat\rho
\end{pmatrix} \ .
\end{equation}
Any Bogolubov transformation with the same $t$ is actually conjugate to this matrix.
The powers of $M$ are given by
\begin{equation}
M^n=\begin{pmatrix}
A_n&B_n\\
C_n&D_n
\end{pmatrix}=\begin{pmatrix} A&B\\
C&D\end{pmatrix}^n = \begin{pmatrix}
\cosh n \hat \rho&\sinh n\hat\rho\\
\sinh n\hat \rho & \cosh n\hat \rho
\end{pmatrix}\label{eq:powM}
\end{equation}
so that  
\begin{equation}
C_n D_n^{-1} = \tanh n \hat \rho \ . \label{eq:s_param}
\end{equation} 

Similarly, for $|t|<2$, $t=2 \cos\hat \theta$, a simple $M$ with the right characteristic polynomial is given by
\begin{equation}
M= \begin{pmatrix}
A&B\\
C&D
\end{pmatrix}= \begin{pmatrix}\exp(i\hat\theta)&0 \\
0& \exp(-i \hat\theta)
\end{pmatrix}
\end{equation}
so in this case the oscillator algebras don't really mix.

\section{Entropy growth I: tracing over anti-particles}
\label{sec:Ent1}

Our goal is to analyze the entanglement entropy of a squeezed state as described in equation 
Eq.~\eqref{eq:squeeze}. This is,  we want to understand the entanglement entropy of a state of the form
\begin{equation}
|\Omega\rangle\propto \exp(\alpha a^\dagger b^\dagger) \vac
\end{equation}
where $\alpha$ is a  complex number.
To do that, we need to choose a factorization into a product of two Hilbert spaces. 
Since in this setup it is natural to separate the Hilbert space according to the oscillator algebras of $a, b$ separately, 
here we will trace over the $b$ oscillators.
This is what one would usually do in the case of black hole physics. 
Then we will compute the growth of the entropy as a function of time according to the iteration in Eq.~\eqref{eq:iter}.

The state $|\Omega\rangle$ needs to be normalized. We will consider normalized states $\ket{n,m}$ for the two sets of oscillators. In terms of these, the states $a^{\dagger k} b^{\dagger k} \vac = k! \ket{k,k}$ have norm
$k!$.  With these we find that 
\begin{equation}
\exp( \alpha a^\dagger b^\dagger)  \vac = \sum \frac{\alpha^k}{k!}a^{\dagger k} b^{\dagger k} \vac
 = \sum \alpha^k \ket{k,k} \ .
\end{equation}
\rmv{As part of the exponential defining the squeezed state, the coefficient in front of the $\ket{k,k}$ state is $\alpha^k$.}
The norm of the state is therefore 
\begin{equation}
\label{eq:norm1}
|\exp( \alpha a^\dagger b^\dagger)  \vac |^2 = \sum_k |\alpha|^{2k} = \frac 1{1-|\alpha|^2}
\end{equation}
and it is finite only if $|\alpha|<1$.

When we trace over the $b$ degrees of freedom we get\rmv{an effective}
\keep{a reduced} density matrix for the $a$ degrees of freedom given by
\begin{equation}
\rho_{a} = (1-|\alpha|^2) \sum |\alpha|^{2k} \ket k \bra k\ . \label{eq:thermaldm}
\end{equation}
This is the same as a thermal density matrix for a harmonic oscillator 
$a, a^\dagger$ with the Boltzmann factor identified as  $\exp(-\beta \hbar \omega) =|\alpha|^2$. 
So from the perspective of the $a$ oscillator, the dynamics is pumping in heat.

The entropy of this density matrix is the entanglement entropy of the subsystem of the $a$ oscillators and it is then given by 
\begin{equation}
S=-\log(1-x) - \frac x{x-1}\log x\label{eq:thermalent}
\end{equation}
with $x=|\alpha|^2$.
Notice that this entropy is independent of the phase of $\alpha$.

We now apply this result to the squeezed state that arises from the iteration of 
our unitary dynamics in the special case of Eq.~\eqref{eq:powM}, which results in the identification
\begin{equation}
|\alpha_n |= \tanh n \hat \rho 
\end{equation}
so that for large $n$ we have
\begin{equation}
x=\tanh^2 n\hat \rho=1-\frac 1{\cosh (n\hat \rho)^2} \to 1-\frac 4 {\exp(2 n\hat \rho)}
\end{equation}
and we find that 
\begin{equation}
S_n \to \log( \exp(2 n \hat \rho)/4)+O(1)= 2 n \hat \rho +O(1) \ .
\end{equation}
We thus find that the asymptotic entropy growth per unit time is characterized by
\begin{equation}
\frac{\Delta S}{\Delta t} = \frac 1 T\frac{\Delta S}{\Delta n}= \frac{2 \hat \rho}{T}
\end{equation}
where we have inserted the time scale of the periodic trajectory alluded to in the previous section.

This shows that the entropy growth is \rmv{essentially}\keep{asymptotically}
linear as a function of time, and that the growth is controlled by the eigenvalues of the matrix $M$. 
We will later interpret this result in terms of\rmv{the Kolmogorov-Sinai entropy growth}
\keep{Lyapunov exponents}. Our goal is now to show that this result is essentially independent of 
the choice of basis for the harmonic oscillators. The first step is to consider a more general 
matrix as in Eq.~\eqref{eq:param}. 
Since $t= 2\cosh \hat \rho = 2 \cosh \rho \cos\theta$, we immediately find that $\rho\geq \hat \rho$ and that therefore 
the entropy for other choices of initial states, 
related to the vacuum by the action of such a general matrix, 
is larger than that for the special case chosen above. 
We will find that this only affects the $O(1)$ piece, not the leading $n$-dependent piece. 

Since each such more general matrix is conjugate to $M$, we can write\rmv{the following}
\keep{its $n$th power as follows:}
\begin{equation}
\begin{pmatrix} e^{i \theta_n} \cosh\rho_n  & e^{i\phi_n} \sinh\rho_n \\
e^{-i\phi_n} \sinh\rho_n & e^{-i\theta_n}\cosh \rho_n \end{pmatrix}= T M^n T^{-1}
\end{equation}
with 
\begin{equation}
T= \begin{pmatrix} e^{i\gamma_2} \cosh\gamma_1  & e^{i\gamma_3} \sinh \gamma_1 \\
e^{-i\gamma_3} \sinh\gamma_1 & e^{-i\gamma_2} \cosh \gamma_1 \end{pmatrix}\ .
\end{equation}
It follows from a straightforward computation that 
\begin{eqnarray}
\cosh  \rho_n  \exp(i\theta_n)&=& \cosh (n \hat \rho) + \cosh(\gamma_1)\sinh(\gamma_1)(\exp(i \gamma_3-i\gamma_2)-\exp(i\gamma_2-i\gamma_3)) \sinh(n \hat \rho)\nonumber\\
&=& \cosh( n\hat \rho) +2 i  \cosh(\gamma_1)\sinh(\gamma_1) \sin(\gamma_3-\gamma_2) \sinh(n \hat \rho) \ .
\end{eqnarray}
With this information we find that $|\cosh \rho_n|^2$ is modified asymptotically as
\begin{equation}
|\cosh  \rho_n |^2 \simeq \frac{1+ |\sinh(2\gamma_1)\sin(\gamma_3-\gamma_2) |^2 }4\exp(2 n \hat \rho) 
\end{equation}
and one can show that in the formula determining the entropy this only affects the order $O(1)$ terms by 
$\log(1+ |\sinh(2\gamma_1) \sin(\gamma_3-\gamma_2) |^2)$. This is independent of $n$, so we still have the result 
\begin{equation}
\label{eq:S_n}
S_n= 2 n \hat \rho +O(1) \ ,
\end{equation}
independent of the choice of initial state, \keep{since such a choice amounts to another conjugation,
as discussed below Eq.~\eqref{eq:Q}.}
Out of these choices of initial state and evolution operator, 
the entropy is smallest for the special matrix $M$ we chose, but asymptotically this difference in negligible.
We see here that the $n$ at which the linear growth term dominates depends on the size of the 
$\gamma_i$ and $\hat \rho$. In practice, if the $\gamma_i$ and $\hat \rho$ are all $O(1)$, then even for $n < 10$ 
the linear approximation is quite good, justifying our remarks in Sec.~\ref{sec:strobo} that we can ignore
other potentially nearby periodic orbits. If one considers more extreme values, 
the value of $n$ at which we reach the asymptotic behavior above depends
on details.

\section{Entropy growth II: The general case}
\label{sec:eII}

One can imagine that, in general, different factorizations of the Hilbert space will lead to different answers for the entanglement entropy and that this might affect the entropy growth we computed in the previous section. The purpose of this section is to show that this is not 
the case: if one uses a different linear combination of raising operator states to define particles versus antiparticles, we will see that in general they all have the same asymptotic growth of the entropy. We will also generalize the results to include general canonical linear transformations, not only those 
that preserve a $\text{U}(1)$  charge.

So far we have chosen to trace over the $b$ oscillators for a pure state of the form
\begin{equation}
\ket\Omega\propto\exp(\alpha a^\dagger b^\dagger)\vac
\end{equation}
to compute the entanglement entropy growth. The simplicity of this particular choice is that the associated density matrix for the $a$ oscillators is already diagonal in the number basis when we trace over the $b$ oscillators. We will now study a more general case where this diagonal form is not readily available. 

The idea here is to rotate the oscillator algebra as follows
\begin{equation}
\begin{pmatrix}
a^\dagger\\
b^\dagger
\end{pmatrix}= \begin{pmatrix} \cos\theta & \sin \theta\\-\sin\theta & \cos \theta \end{pmatrix}\begin{pmatrix}
a_{r}^\dagger\\
b_{r}^\dagger
\end{pmatrix} \ .\label{eq:rot1}
\end{equation}
Here the subscript $r$ is reminding us that the basis is rotated.
Such a rotation is also a Bogolubov transformation that preserves the ladder operator algebra, but 
it does not correspond to an evolution like the one we have  considered before, 
since it violates the $\text{U}(1)$ symmetry, i.e., it does not preserve 
the Noether charge $Q = a^{\dagger}a - b^{\dagger}b$ (except for $\theta = 0$). 
The vacuum $\vac$ does not change under such a rotation, but the new way of writing the state is more complicated.
It can be thought of as a redefinition of particles and anti-particles at a given time, 
or a different coarse graining of the system into object and environment. 
The underlying state can be written
\begin{equation}
	\ket\Omega = C\exp \big{[} \alpha (-\cos\theta\sin\theta a_{r}^{\dagger 2}  + (\cos^{2}\theta - \sin^{2}\theta) a_{r}^{\dagger} b_{r}^{\dagger} + \cos\theta\sin\theta b_{r}^{\dagger 2}) \big{]}\vac \ . \label{eq:sqrot}
\end{equation}
We now want to trace over the $b_r$ oscillators and compute the density matrix for the $a_r$ oscillators. 
We can also consider a more general $\text{SU}(2)$ rotation for equation \eqref{eq:rot1}. This will modify various relative phases in equation \eqref{eq:sqrot}, but the structural form will be very similar, since the additional phases can be eliminated by changing the phase of the $a^\dagger, b^\dagger$ and the $a_{r}^\dagger, b_{r}^\dagger$ operators. 
So, we can assume equation \eqref{eq:rot1} without loss of generality.

A convenient way to compute the reduced density operator is to use the 
holomorphic representation of the oscillator Hilbert space (also called the Bargmann representation). 
Here $a^\dagger \simeq z$ and $a\simeq \partial_z$.
We can arrive at this representation by noticing that for any entire function $\psi(z)$ there is a corresponding state $\ket\psi= \psi(a^{\dagger})\vac$ and that the annihilation operator acts as $\partial/\partial z$, i.e., $a\ket\psi = \psi'(a^{\dagger})\vac$. For two states $\ket\phi$ and $\ket\psi$ we can write their inner product as
\beq
\label{eq:inprod}
	\langle \phi | \psi\rangle  = \frac1{\pi} \int \phi(\bar{z}) \psi(z) e^{-z\bar z} \ dzd\bar z.
\eeq
\keep{See Appendix \ref{ap:entropy} for further details.}
Thus we may represent our state by
\begin{align}
	\ket\Omega \rightarrow \Omega(z,w) &= C' \exp \big{[} \alpha (-\cos\theta\sin\theta z^{2}  + (\cos^{2}\theta - \sin^{2}\theta) zw + \cos\theta\sin\theta w^{2}) \big{]} 
\end{align}
and similarly
\beq
	\bra \Omega \rightarrow \bar\Omega(\bar z,\bar w) = \bar C' \exp \big{[} \bar{\alpha} (-\cos\theta\sin\theta \bar z^{2}  + (\cos^{2}\theta - \sin^{2}\theta) \bar z\bar w + \cos\theta\sin\theta \bar w^{2}) \big{]}.
\eeq
Then we can implement the trace over the $b$ oscillators via the integral
\beq
	\rho_a = \Tr_b \ket \Omega \bra \Omega \rightarrow \rho_{a}(z,\bar z) = \frac1{\pi} \int \Omega(z,w) \bar\Omega(\bar z, \bar w) e^{-w\bar w} dw d\bar w,
\eeq
with an abuse of notation by writing the holomorphic representation of $\rho_{a}$ as $\rho_{a}(z,\bar z)$.
Since the integrand is the exponential of a quadratic polynomial, we can perform the integral. 
For simplicity of notation let us assume that $ \alpha \in \mathbb{R}$, which can be achieved by changing the phases of the variables $z, w$.

\begin{figure}[ht]
\begin{center}
\includegraphics[width=4in]{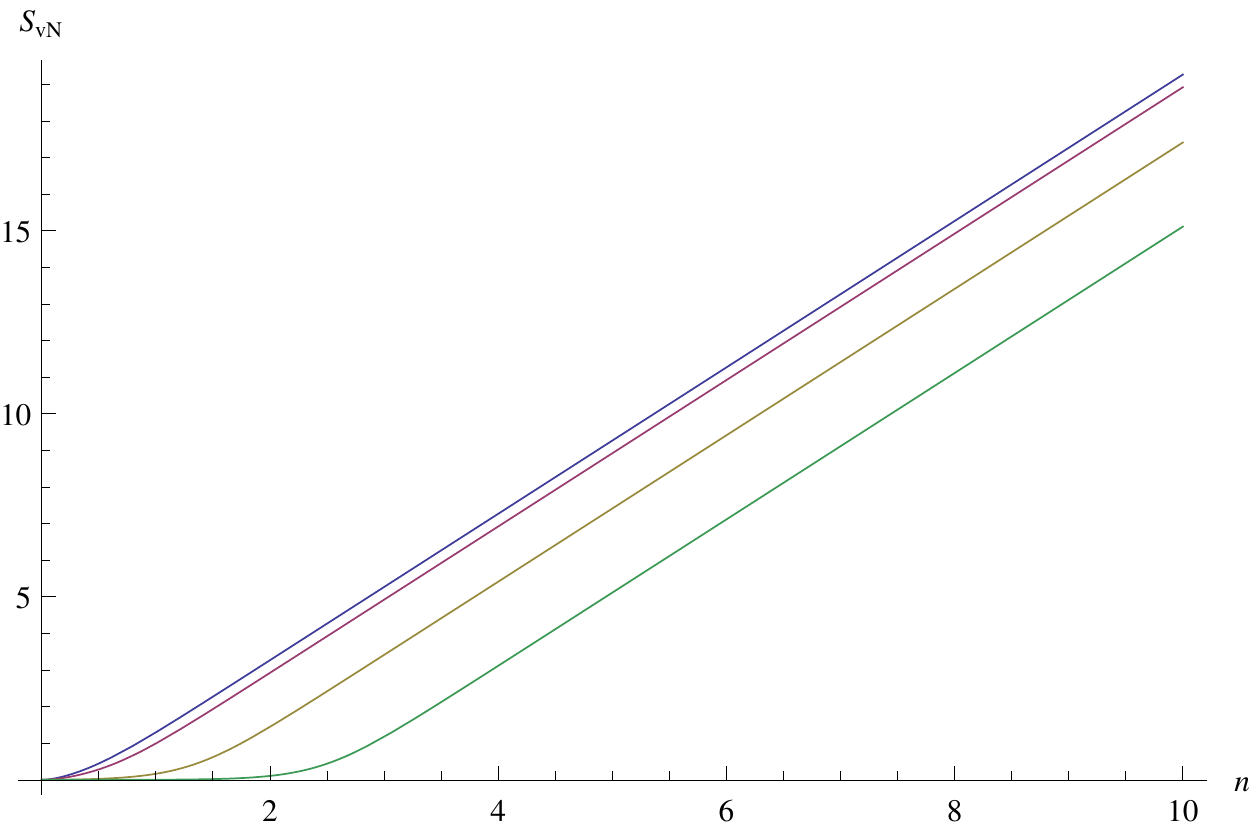}
\end{center}
\caption{The von Neumann entropy of $\rho_{a}$, as given by Eq.~\eqref{eq:vNEnt}, 
with $\alpha_n = \tanh n$, as a function 
of $n$, for $\theta = 0, (0.5)\pi/4, (0.9) \pi/4, (0.99)\pi/4$ (topmost to bottom-most curve) . }
\label{fig:SvN_n_theta_1}
\end{figure}

We get
\beq
\label{eq:rhoa}
	\rho_{a}(z,\bar z) = \frac{1- \alpha^{2} }{\sqrt{1 - \alpha^2 \sin^2 2\theta}} \exp\left[\alpha \frac{(\alpha^2-1)\sin(2\theta) (z^2 +\bar{z}^2) + 
		2\alpha \cos^2 (2\theta) z\bar{z} }{2(1-\alpha^2 \sin^2 2\theta) } \right],
\eeq
where we used $|C'|^{2} = 1- \alpha^{2}$ (see Eq.~\eqref{eq:norm1}).
To get the $(m,n)$ matrix element of $\rho_a$ we simply read off the coefficient of $z^m \bar z^n$ in the above, times
$\sqrt{m! n!}$. 
It is useful to rewrite $\rho_{a}$ as 
\beq
\label{eq:rhoa2}
	\rho_{a}(z,\bar z) = B_{0} e^{B_{1}(z^{2} + \bar z^{2}) + B_{2} z \bar z} \ ,
\eeq 
where 
\begin{align}
	B_0 &= \frac{1- \alpha^{2} }{\sqrt{1 - \alpha^2 \sin^2 2\theta}} \ , \\
	B_1 &=  \frac{\alpha (\alpha^2-1)\sin 2\theta}{2(1-\alpha^2 \sin^2 2\theta)}  \ , \\
	B_2 &= \frac{\alpha^2 \cos^2 2\theta}{1-\alpha^2 \sin^2 2\theta}  \ . 
\end{align}
This is the same form of density matrix as that obtained in \cite{2000JPhA...33.8139B}, where they were able to compute
the entanglement entropy\rmv{ by finding an ansatz for the corresponding density operator that is 
a pure exponential of $a$ and $a^\dagger$ operators}. 
We reproduce their computation in detail in Appendix \ref{ap:entropy} and find
\beq
\label{eq:vNEnt}
S = -\Tr (\rho_{a} \log \rho_{a}) = (\chi+ \tfrac12)\log (\chi+ \tfrac12) - (\chi- \tfrac12)\log (\chi- \tfrac12) \ ,
\eeq
where 
\beq
	\chi = \sqrt{ \frac{\alpha^2 \cos^2 2\theta }{(1-\alpha^2)^{2}} + \frac14} \ . \label{eq:chidef}
\eeq
\rmv{We explain these computations in detail in the appendix \ref{ap:entropy}.}
What is important for us is that for large times we have that $\alpha_n \to 1$, so 
\keep{substituting $\alpha_n$ for $\alpha$ in Eq.~\eqref{eq:chidef}}, the resulting value of $\chi_n$ 
becomes large and is dominated by the singularity in the denominator inside the square root. 
For large $\chi$, we can expand the entropy as follows
\begin{equation}
\label{eq:S_n2}
S \approx (\chi+\tfrac 12) \log (\chi)-(\chi-\tfrac 12) \log(\chi) +O(1) \approx \log(\chi)+O(1)
\end{equation}
and\rmv{$\log(\chi) = \log( 1/(1-\alpha_n^2))+O(1)$}
\keep{$\log(\chi) \approx \log( 1/(\alpha_n^2-1))+O(1)$}. Substituting $|\alpha_n|=\tanh n \hat \rho$ 
we get the same asymptotic answer as 
before, but the terms of order one now act to lower the entropy rather than to raise it. 
See Fig.~\ref{fig:SvN_n_theta_1} for a plot of the entropy as a function of $n$ for various choices of 
$\theta$.

It should be noted that this reasoning is correct so long as $|A_2|\neq 0$, that is, $\cos 2\theta \neq 0$. For that very special case, the coefficient of 
$a^\dagger b^\dagger$ in the squeezed state in Eq.~\eqref{eq:sqrot} vanishes, 
and the state is a product state between the two Hilbert spaces. 
In that case the entropy vanishes and $\chi=\tfrac 12$. 
Thus, there is one special selection of coarse graining that does not produce entanglement entropy.

The other case we have not treated so far is when instead of having a 
complex eigenvalue for the canonical transformation (and its three images), 
we have two different real eigenvalues and their inverses. In such a case, 
to each pair of eigenvalue we can associate a $q, p$ pair in phase space, and 
ladder operators $c_{i}^\dagger,c_i$, as before.
The Bogolubov transformations will mix the $c,c^\dagger$ operators for each $i$ independent from each other. Such a transformation can be parametrized as 
\begin{equation}
c= (\cosh \rho) c_{\text{New}}- (\sinh \rho) c_{\text{New}}^\dagger
\end{equation}  
and this leads us to a squeezed state given by
\begin{equation}
\ket 0 _0\propto \exp\left(  \frac 12 (\tanh \rho_1) c_1^{\dagger 2}+ \frac 12 (\tanh \rho_2) c_2^{\dagger 2} \right) \ .
\end{equation}
This is a product state, and there is apparently no entropy production. 
In the holomorphic basis, this corresponds to a state
\begin{equation}
\ket 0_0 \propto \exp\left(  \frac 12 (\tanh \rho_1) z^2+ \frac 12 (\tanh \rho_2) w^2\right)
\end{equation}
and to eigenvalues $\exp(\pm \rho_{1,2})$ for the different directions in phase space.

Upon iteration, we would get
\begin{equation}
\ket 0_0 \propto \exp \left(  \frac 12 (\tanh n\rho_1) z^2+ \frac 12 (\tanh n \rho_2) w ^2\right)=
	\exp\left(\frac {s_1}2 z^2+\frac {s_2}2 w^2\right) \ .
\end{equation}
What is important to notice is that so long as $\rho_{1,2}$ are non-zero, the large iteration limit produces coefficients that tend to $1/2$, and we have introduced $s_1,s_2$ in order to write compact expressions later.

Just like before, we can rotate the algebra and ask what 
the dependence of the entropy growth on the choice of rotation parameter is. 
The rotations that are available to us given $c_{1,2}^\dagger$ belong to $\text{SU}(2)$. We are interested in the large iteration behavior. 
However, notice that if we do a linear transformation between $c_1^\dagger,c_2^\dagger$ with real parameters, we 
get a cancellation of the cross term in the squeezed state, producing a state that is still factorized. 
Thus, it is more convenient to consider a complex transformation for illustration purposes:
\begin{eqnarray}
z= \cos(\theta) z_1+ i \sin(\theta) w_1\\
w= i \sin(\theta) z_1 + \cos(\theta) w_1
\end{eqnarray}
Alternatively, one can take $\rho_2\to -\rho_2$ and use a real rotation between $z,w$ to go to the new basis. Integrating out the new variable $w_1$ again produces a density matrix of Gaussian form.
The corresponding squeezed state in the new basis is
\begin{equation}
\ket 0_0 \propto \exp \left( \frac{s_1}{2}  (\cos(\theta) z_1+ i \sin(\theta) w_1)^2+\frac {s_2}{2} (i \sin(\theta) z_1 + \cos(\theta) w_1)^2 \right)
\end{equation}
Call the quadratic form in the exponent $Q(z_1,w_1)$. To obtain the density matrix for $z_1$, 
we take the partial trace over the $w_1$ modes. This is given by
\begin{equation}
\rho_a(z_1, \bar z_1)\simeq \int d^2 w_1 \exp[\bar Q(\bar z_1, \bar w_1)+Q(z_1,w_1)-w_1\bar w_1] \ .
\end{equation}
The quantity in the exponential is a gaussian with a shift in $w_1,\bar w_1$. When we integrate over $w_1,\bar w_1$, we replace them by the values at the critical point of the Gaussian for fixed $z_1,\bar z_1$. The final answer for this quadratic form is not particularly illuminating. Instead, we quote directly the value of $\chi^2$ which is
\begin{equation}
\chi^2= \frac{(1+s_1 s_2)^2 -(s_1+s_2)^2 \cos^2(2\theta)}{4(1-s_1^2)(1-s_2^2)} \ .
\end{equation}
Notice that this is symmetric in the exchange $s_1\leftrightarrow s_2$, as it should be: 
the entanglement entropy of subsystem $a$ is the same as the entanglement entropy of subsystem $b$.
Inserting $s_1=\tanh(n \rho_1)$ and $s_2=\tanh(n\rho_2)$ we get that 
\begin{equation}
\chi^2= \frac 1{16} [3+\cosh(2 n (\rho_1+\rho_2))-2 \cos(4\theta) \sinh(n(\rho_1+\rho_2)^2) ] \ ,
\end{equation}
and as a consistency check we find that $\chi^2|_{\theta=0}=1/4$ (for which no entropy is generated).
Taking the large iteration limit, we see that the entropy $S\simeq \log(\chi) $ generically grows as 
\begin{equation}
S= \log\exp(n(\rho_1+\rho_2))+ O(1)\simeq n (\rho_1+\rho_2)
\end{equation}
except on a set of measure zero (where $\cos(4\theta)=1$).
Notice also that  when $\rho_1=\rho_2=\hat \rho$ we recover the results of the previous section. 
In general, the entropy grows as the sum of the logarithm of the two large eigenvalues of the linear symplectic transformation.
This result also includes the case where one pair of eigenvalues is unitary and the other is real. 
The unitary (stable) eigenvalue pair does not contribute to the entropy because in that case $s_2=0$.

As a further comment, we should note that all of the calculations work essentially the same way if we start initially with a coherent state, rather than just a vacuum. This is because  a coherent state equation $a \ket{f}=f\ket{f}$ can be thought of as defining a new set of ladder operators via a shift 
$\tilde a=a-f$ and $\tilde a^\dagger= a^\dagger -f^*$. 
This is an automorphism of the algebra of raising and lowering operators (it preserves the commutation relations and the adjoint operation), so it can be used to show that any such coherent state is like a vacuum, and any ``shifted" squeezed state is a squeezed state with respect to a new set of shifted operators. 
Essentially, if we extend the linear problem of Bogolubov rotations to allow these translations, we get an affine group. As such, any element can be composed of a ``rotation"  (a symplectic matrix) plus translation and the translation part can be moved from multiplying on the right to the left without affecting the rotation. This is, if we think of $f$ as a translate of the vacuum
\begin{equation}
\ket{f}\simeq T(f) \vac
\end{equation}
by a unitary operator $T(f)$, and then apply a Bogolubov rotation $R(\theta)$ to it (as an active transformation), we find that since
\begin{equation}
R(\theta) T(f) \simeq T(f') R(\theta) 
\end{equation}
we get that 
\begin{equation}
R(\theta)T(f) \vac \simeq T(f') R(\theta) \vac
\end{equation}
and the effects of $T(f')$ can be undone on the algebra of operators of the target by a passive transformation (it acts on the definition of the algebra, not the states). These translations furthermore preserve the splitting of the Hilbert space that we have described so far (for any choice), so they do not entangle the Hilbert spaces further 
and when we compute the entanglement entropy, \keep{it doesn't affect the answer}.

\section{Comparison with the classical dynamics}
\label{sec:class}

Consider a general chaotic Hamiltonian dynamical system with finitely many degrees of freedom, 
corresponding to canonical pairs $q_i, p_i$, and with a time independent Hamiltonian. 
Assume that we have a periodic trajectory with period $T$, characterized  by $q^{(0)}_i(t), p_i^{(0)}(t)$ and consider the evolution of infinitesimal deviations, 
$\delta q_i(t), \delta p_i(t)$, from that trajectory.
The equations of motion are given by
\begin{eqnarray}
\dot q_i &=& \frac{\partial H}{\partial p_i}\\
\dot p_i &=& -\frac{\partial H}{\partial q_i} \ .
\end{eqnarray} 
Using $q_i= q_i^{(0)}+\delta q_i$, $p_i= p_i^{(0)}+\delta p_i$, we can Taylor expand
\begin{equation}
\dot q_i = \dot q_i^{(0)}+ \delta \dot q_i =  \frac{\partial H}{\partial p_i}{\bigg\vert}_{q^{(0)}, p^{(0)}} 
	+\sum_j\frac{\partial H}{\partial p_i \partial s_j}{\bigg\vert}_{q^{(0)}, p^{(0)}} \delta s_j+O(\delta s^2)
\end{equation}
and similarly for $\dot p$. Here the sum $s_j$ is over all the canonical variables. Since $q^{(0)}$ is a solution of the equations of motion, we find that the evolution of the $\delta s_j$ is given by a first order set of coupled homogeneous linear differential equations. These have time dependent coefficients characterized by the quantities 
\begin{equation}
S_{ij} (t) \simeq \pm \frac{\partial H}{\partial s_i \partial s_j}{\bigg\vert}_{q^{(0)}, p^{(0)}}
\end{equation}
which are also periodic with period $T$.
There is a monodromy matrix $\Phi(T)$ associated with this problem, a Floquet problem, that expresses any solution after time $T$ in terms of known solutions with $t<T$, in the form $u(t+T) = \Phi(T)\cdot u(t)$. 
\keep{Here $u(t)$ is a vector of linearly independent solutions to the equations of motion, and $\Phi(T)$ is 
a representation of the Floquet operator (see, e.g., \cite{Jose1998} for a review of this formalism).}
This can be iterated, so that $u(t+nT)= \Phi(T)^n \cdot u(t)$.
The monodromy matrix $\Phi(T)$ plays the role of $M$ (the evolution matrix discussed in section \ref{sec:strobo}). 
One of the eigenvalues of $\Phi(T)$ is $1$, corresponding to the the eigenvector 
\keep{representing} the infinitesimal deviation $\delta s_i = \dot s_i^{(0)} \delta t$. 
This just corresponds to changing $t\to t+\delta t$. 

As above, we assume that we are in a generic case, so that 
we can decompose our vector space along eigenvectors of $\Phi(T)$ and 
 only keep the span of the vectors that correspond to eigenvalues different than $1$. 
 This is the space which we assumed\rmv{has only dimension $2$ (in the sense of canonical pairs)}
 \keep{is four-dimensional} in the previous sections.

One can do something similar on Hamiltonian systems with a periodic driving and with a periodic trajectory, 
which are associated\rmv{to} \keep{with} a time-dependent Hamiltonian $H(t)$. 
Such systems can be extended from $N$ to $N+1$ degrees of freedom, where one takes 
$H_{\text{New}}= p_s+H(s)$ with $(s,p_s)$ the new pair of canonical variables. 
The equations of motion of $s$ are $\dot s=1$, so $s$ is essentially time and now any such 
time dependence is due to solving the evolution of a time independent system. 
These systems are said to have $N+1/2$ 
degrees of freedom \cite{Zaslavsky2005,Zaslavsky2007}.
They have the advantage that eigenvalue that is equal to unity can be associated\rmv{to} \keep{with} the pair 
$(s,p_s)$ itself, so the associated monodromy matrix $\Phi(T)$ can be thought of as non-degenerate 
and even-dimensional from the get go.

We now consider the Lyapunov exponents for this system. 
Let us assume we are in the generic case, and 
write the eigenvalues of $\Phi(T)$ in groups of four, $\lambda, \bar\lambda, \lambda^{-1}, \bar{\lambda}^{-1}$, 
all distinct, as above, with $\lambda= \exp(\rho+i \beta)$. We take $\rho > 0$ since 
that is the case in which we are most interested.
Let us further work in a (possibly complex) basis that diagonalizes $\Phi(T)$, so that the Jacobian 
matrix of the dynamics in this basis is simply 
$J = \text{diag}(e^{\rho+i \beta}, e^{\rho-i \beta}, e^{-\rho- i \beta}, e^{-\rho+i \beta})$.\footnote{
We are effectively using covariant Lyapunov vectors 
\cite{2007PhRvL..99m0601G, 2013JPhA...46y4005G} (see also \cite{Cvitanovic2012} Ch. 6).} 
The Lyapunov exponents are the logarithms of the eigenvalues of the matrix 
\beq
L = \lim_{n\rightarrow\infty} \left( (J^n)^\dagger J^n  \right)^{1/2n} = \text{diag}(e^{\rho}, e^{\rho}, e^{-\rho}, e^{-\rho}) \ ,
\eeq
so for 
each such group one has the Lyapunov exponents $(\rho,\rho, -\rho, -\rho)$. 
The exponents for the continuous-time system have to be scaled by $T$, so 
this indicates the rate of separation of nearby points grows exponentially in time as $\exp(\rho t/T)$.

For sufficiently uniform chaotic systems on domains with finite measure  
the Kolmogorov-Sinai (KS) entropy is equal to the sum of the positive
Lyapunov exponents $\kappa_i$ (times their degeneracy $d_i$):
\begin{equation}
h_{\text{KS}}=\sum_{\kappa_i > 0} d_i \kappa_i   \ ,
\end{equation}
by Pesin's theorem.\footnote{For more general systems, the relationship is given by the
Ruelle inequality \cite{MR516310, Ruelle1989, Katok1995}.} 
This is the sum of all expanding Lyapunov exponents. 
Roughly speaking, the KS entropy expresses the rate at which we lose information about the 
initial conditions due to the chaotic dynamics.
One starts by subdividing the set of possible initial conditions into a 
 coarse-grained set of small volume elements, or cells. 
 The KS entropy computes how many such cells are covered by our 
 initial volume when we evolve it to asymptotically large times and take the size scale of the coarse-graining to zero, 
see, e.g., \cite{Zaslavsky2005, Katok1995}.

We can use the same intuition in our case, even though we are considering dynamics in 
a tangent space approximation, and so do not have a finite measure space.\footnote{There are 
various rigorous definitions of local entropy and entropy for non-compact or non-finite measure spaces, 
e.g., \cite{Ledrappier1985, ETS:7891821}, but here we do not attempt to use these, and so essentially 
take Eq.~\eqref{eq:S_c} as our definition of $S_{\text{c}}$. }
That is, there is a classical entropy $S_{\text{c}}(t)$ associated with some coarse-graining, that 
in the limit of small cell size and late times grows as 
\begin{equation}
\label{eq:S_c}
S_{\text{c}}(t) \approx h_{\text{KS}} t = \left(\sum_{\kappa_i > 0} d_i \kappa_i\right) t \Rightarrow S_{\text{c}}(nT) \approx 2n\rho \ ,
\end{equation}
which is exactly what we found in the quantum calculations, see Eqs.~\eqref{eq:S_n} and \eqref{eq:S_n2}.

We thus see that the asymptotic growth rate of the entanglement entropy 
is\rmv{identically} equal to the sum of the positive Lyapunov exponents, 
implying the asymptotic convergence of the classical entropy and entanglement entropy.

\section{Discussion}


In this paper we studied entanglement entropy growth under a Bogolubov transformation that entangled two degrees of 
freedom and their canonical conjugates. To understand entanglement entropy  we traced over half of the degrees of freedom. The rate growth of the entanglement entropy ended up being asymptotically equal to 
the sum of the two positive Lyapunov exponents of the system. This was true for all factorizations except on sets of measure zero.

We have seen that in the simple system we studied here there is a strong 
relationship between a notion of classical dynamical entropy and the growth of 
entanglement entropy. Namely, the former bounds the latter and they are almost always 
asymptotically equal. 
This is an indication that the classical dynamics, 
as encoded by dynamical systems quantities like the entropy, contains significant 
information about the associated quantum dynamics and the rates of scrambling for such systems, 
in line with many previous results on quantum dynamical systems.
We note here though, that we considered almost-classical initial conditions. 
It would be interesting to analyze the dynamics of more intrinsically quantum states, like the superposition of 
widely separated coherent states.

Our results can be generalized 
to more degrees of freedom, like the general linearized dynamics near periodic orbits for higher dimensional chaotic Hamiltonian systems. 
In the simplest cases, the degrees of freedom can be treated 
as independent and
we expect that the growth of entanglement entropy will 
again asymptotically converge to the classical entropy.
For example, if we partition the set of coordinates into two sets, one of $k$ coordinates and one of $n-k\geq k$ coordinates 
(where by coordinates here we mean canonical pairs
written as ladder operator pairs), we expect the entanglement entropy for the factorization should grow asymptotically 
like the sum over the $2k$ largest positive Lyapunov exponents, except on sets of measure zero for the possible factorizations.
This comes from the additivity of entanglement entropy for independent systems. 

Note that in these systems the Lyapunov exponents are paired: for every positive Lyapunov exponent $\kappa$, there is a corresponding negative one $-\kappa$, due to the conservation of the symplectic form. The maximum entropy growth would then be achieved when $k$ is as close as possible to $n-k$, and we would get that the entropy growth would be the  sum of all positive Lyapunov exponents. This is the same way that classical entropy (as described by Pesin's formula) is supposed to behave for general dynamical systems. 
However, the details of the general case of $N$ degrees of freedom remain for future work.
	
There are other definitions of entropy in quantum dynamical systems, e.g., \cite{1987CMaPh.112..691C, 1994LMaPh..32...75A}, 
and it would be interesting to understand the relationship between these and the entanglement entropy
studied here, for the kinds of systems we are considering. 
Additionally, the dynamics studied here was confined to a linearized analysis about some periodic 
orbit. To extend these results beyond this, we will have to confront 
the effects of folding and the highly non-trivial topology of 
chaotic domains in phase space \cite{Zaslavsky2005, Zaslavsky2007, Cvitanovic2012}.

\acknowledgments
We thank M. Srednicki for many discussions, and 
CA thanks F. Denef, M. Fannes, R. Monten, and Y. Lemonik for helpful discussions.
Work of DB supported in part by the U.S. Department of Energy under grant DE-SC0011702. 
The research leading to these results has received funding from the European Research Council under the European Community's Seventh Framework Programme (FP7/2007-2013) / ERC grant agreement no. [247252].
CA is supported in part by a grant from the John Templeton Foundation and in part by the U.S. Department of Energy under grant DE-FG02-92-ER40699. 
The opinions expressed in this publication are those of the authors and do not necessarily reflect the views of the John Templeton Foundation.

\appendix

\section{General Bogolubov transformation preserving charge}
\label{ap:bog}

We consider a phase space with two oscillators, whose ladder operator algebra is given by the operators $a^\dagger, a$ and $b^\dagger, b$. 
We will assume that $a^\dagger$ has charge $+1$ and $b^\dagger$ has charge $-1$ and that a discrete linear dynamics (generated by a unitary operator)  on this system preserves the $\text{U}(1)$ charge. The commutation relations are
\begin{eqnarray}
\! [a,a^\dagger]&=& 1 = [b, b^\dagger] \label{eq:osc}\\
\! [a^\dagger, b^\dagger]&=&0=[a^\dagger, b]\quad \hbox{etc.}\label{eq:orth}
\end{eqnarray}

The most general transformation that is linear and unitary and preserves the algebra 
is given by a Bogoliubov transformation to oscillators $\tilde a^\dagger, \tilde a$ and $\tilde b^\dagger, \tilde b$. The most general such linear transformation compatible with the $\text{U}(1)$ symmetry is given by
\begin{equation}
\begin{pmatrix}
\tilde a^\dagger\\
\tilde b
\end{pmatrix}=\begin{pmatrix} A& B\\
C&D
\end{pmatrix}\begin{pmatrix}
 a^\dagger\\
 b
\end{pmatrix} \ .
\end{equation}
Taking adjoints we have that 
\begin{equation}
\begin{pmatrix}
\tilde a\\
\tilde b^\dagger
\end{pmatrix}=\begin{pmatrix} A^*& B^*\\
C^*&D^*
\end{pmatrix}\begin{pmatrix}
 a\\
 b^\dagger
\end{pmatrix}
\end{equation}
and unitarity implies that the commutation relations are not changed between the $a,b$ and $\tilde a, \tilde b$ operators.
We need to verify those of equation \eqref{eq:osc} and the one on the left of equation \eqref{eq:orth}.
Using the commutation relations we find that
\begin{equation}
[\tilde a, \tilde a^\dagger]= A A^* [a,a^\dagger]+BB^* [b^\dagger, b]= |A|^2-|B|^2=1 \ .
\end{equation}
We can thus parametrize the solutions of this equation 
as $ A= \cosh (\rho) \exp(i\theta+i\beta)$, $B=\sinh(\rho) \exp(i\phi+i\beta)$
where we have added a redundancy of the description with the angle $\beta$.
We also get that
\begin{equation}
[\tilde a^\dagger, \tilde b^\dagger]= A C^*[a^\dagger, a]+ B D^*[b, b^\dagger]= BD^* -A C^*=0 \ .
\end{equation}
Thus 
\begin{equation}
\frac{D}{C}= \frac {A^*}{B^*} = \frac{\exp(-i \theta ) \cosh \rho }{ \exp(-i\phi) \sinh \rho} \ .
\end{equation}
Finally, we also get that $|D|^2-|C|^2=1$, so $D,C$ differ from $A^*, B^*$ by a phase. We choose the 
phase so that it is easily expressible in terms of  $\beta$, so that we get
\begin{equation}
\label{eq:paramapp}
\begin{pmatrix} A& B\\
C&D
\end{pmatrix}= \exp(i\beta)  \begin{pmatrix} e^{i \theta} \cosh\rho & e^{i\phi}\sinh\rho \\
 e^{-i\phi}\sinh\rho & e^{-i \theta}\cosh \rho \end{pmatrix} \ .
\end{equation}


\section{Computing entropy for Gaussian density matrices}
\label{ap:entropy}

Consider the holomorphic quantization of the harmonic oscillator, also called the Bargmann representation, 
see, e.g., \cite{2006JPhA...39R..65V, 2000JPhA...33.8139B} for reviews.
An arbitrary state vector $\ket{f}$ in the Hilbert space of the harmonic oscillator 
is represented by a holomorphic function 
$f(z)$ and the inner product is given by
\begin{equation}
\braket{f}{g}=\frac 1{2\pi i} \int_{\mathbb{C}} d^2 z \exp(-z\bar z) f^*(\bar z)g(z) \ .
\end{equation}
Here we define the integral by taking $z= x+iy$, $\bar z=x-iy$, and the measure $d^2z = d\bar z\wedge d z= 2i dx\wedge dy $. 
The limits of integration for $x,y$ are from $-\infty$ to $\infty$. 
For example, take the states $\ket{z^n}$ represented by  $z^n$.
With this inner product we have that 
\begin{equation}
\langle z^m \vert z^n \rangle = \frac 1{2\pi i} \int d^2 z \exp(-z\bar z) \bar z^m z^n = m! \delta_{m,n}
\end{equation}
The raising and lowering operators of the harmonic oscillator act as multiplication by $z$ and derivatives with respect to $z$, respectively. 
Orthonormal energy eigenstates $|n\rangle$ are represented by $z^n / \sqrt{n!}$.
A general bounded linear operator $\hat O$ can be written as a formal sum  of ket-bra combinations 
and thus represented by an integral kernel 
$\hat O\to O(z, \bar w)$ whose action on $|f\rangle$ is defined by
\begin{equation}
\hat O\ket{f}\to \frac 1{2\pi i} \int d^2 w O(z,\bar w)\exp(-w\bar w) f(w) \ ,
\end{equation}
which produces a holomorphic function of $z$. The trace of an operator is 
given by
\beq
	\Tr \hat{O} = \frac 1{2\pi i} \int d^2 z \exp(-z\bar z) O(z,\bar z) \ .
\eeq

We assume that $O(z,\bar w)$ is holomorphic in 
$z$ and $\bar w$, and hence can be written
\beq
	O(z, \bar w) = \sum_{n, m =0}^\infty O_{nm} z^n \bar{w}^m \ .
\eeq
The coefficients of this expansion are the matrix elements of $\hat{O}$ in the 
energy eigenbasis:
\beq
	O_{nm} = \langle n \vert \hat{O} \vert m \rangle \ .
\eeq
Determining the kernel of a given operator, and vice versa, is thus straightforward in principle, but
can be difficult in practice if the matrix elements are not simple functions of $n$ and 
$m$. However, for Gaussian operators the problem is tractable because they can 
be specified by just a few parameters.

We are specifically interested in 
density operators whose kernels have a Gaussian form:
\begin{equation}
\label{eq:stp}
\rho(z, \bar z) = B_0 \exp(B_1 z^2+ B_2 z\bar z+B_1^*\bar z^2) \ ,
\end{equation}
which arose in our Eqs.~\eqref{eq:rhoa} and \eqref{eq:rhoa2} 
(here we treat $z$ and $\bar z$ as independent). 
These density operators represent generalized squeezed states. 
To compute the entropy of these density operators we would like to find 
an exponential form so that we can write down $\log \rho$. 
Their entropy was computed this way in 
\cite{2000JPhA...33.8139B} and we reproduce their computation here,
giving more details of the derivation.
To this end first consider the parametrization 
\begin{equation}
\label{eq:rho1}
\rho(z, \bar z)= \frac 1{\sqrt N}  \exp\left(- \frac 1{2 N} (  -x z^2 +2 y  z\bar z-x^*\bar z^2)+\bar z z\right) \ .
\end{equation}
The parameters here correspond to the expectation values of certain simple operators:
\begin{align}
\label{eq:ev1}
	\langle a^2 \rangle &= \Tr (a^2 \rho) = \frac{1}{2\pi i} \int d^2 z\ e^{-z\bar z} \frac{\partial^2}{\partial z^2}
		\rho(z, \bar{z}) = x \\
\label{eq:ev2}
	\langle a^{\dagger 2} \rangle &= \Tr (a^{\dagger 2} \rho) =  \frac{1}{2\pi i} \int d^2 z\ e^{-z\bar z} z^2
		\rho(z, \bar{z}) = \bar{x} \\
\label{eq:ev3}
	\langle a^{\dagger} a \rangle &= \Tr (a^{\dagger} a \rho) =  \frac{1}{2\pi i} \int d^2 z\ e^{-z\bar z} z \frac{\partial}{\partial z} 
		\rho(z, \bar{z}) = y\ .	
\end{align}
Normalization of $\rho$, $\Tr(\rho)=1$, implies that $N= y^2-|x|^2$, while hermiticity requires $y$ to be real 
and $y > 0$. 
Comparing with equation \eqref{eq:stp} we find that 
\begin{eqnarray}
\frac x{2N} &=& B_1\\
1-\frac y{N}&=& B_2\\
(\sqrt N)^{-1}&=&B_0 
\end{eqnarray}
We can easily solve for $N$, finding that 
\begin{equation}
N= [(B_2-1)^2-4 |B_1|^2]^{-1}
\end{equation}
from which $x,y$ readily follow.
It will be useful below to define the following quantity
\begin{equation}
\label{eq:chi1}
\chi^2 = \left(y-\frac 12\right)^2 -|x|^2= \frac{B_2}{B_0^2} +\frac14 = \frac{ 4 |B_1|^2-(1+B_2)^2}{16 |B_1|^2- 4(1-B_2)^2} \ .
\end{equation}

The von Neumann entropy for a density operator is defined as 
\beq
	S(\rho) = -\Tr (\rho \log\rho ).
\eeq
To compute $\log \rho$ we look for an explicitly 
exponential form for $\rho$ (as an operator, not just its representation). 
The Gaussian form of the density operator representation in 
Eq.~\eqref{eq:rho1} suggests a Gaussian ansatz for the operator itself, i.e.,
\beq
\label{eq:rho2}
	\rho = C \exp\left[ -\frac{1}{2}\left( A(aa^\dagger + a^\dagger a) + B a^2 + \bar{B} a^{\dagger 2}  \right) \right] \ .
\eeq
We take $A \in \mathbb{R}$ and $A \geq |B|$ to ensure convergence.
Our starting assumption is that an operator of this form does indeed correspond to 
Eq.~\eqref{eq:rho1}. If so, then it is completely characterized by 
the expectation values of the operators $a^\dagger a$, $a^{\dagger 2}$ and $a^2$, 
so we will compute these and then perform the matching via Eqs.~\eqref{eq:ev1} - \eqref{eq:ev3}.

To compute these expectation values for the density operator in Eq.~\eqref{eq:rho2}, 
first notice that the form of the exponent can be simplified, 
since it is of the form of a standard thermal density operator after a Bogoliubov transfomation.
We can use the inverse transformation to bring the exponent in Eq.~\eqref{eq:rho2} to the standard form.
Specifically, we write an inverse Bogoliubov transformation as
\beq
	\begin{pmatrix}
	a \\ a^\dagger
	\end{pmatrix} =
	\begin{pmatrix}
	\cosh \theta & -e^{i\phi} \sinh\theta \\ -e^{-i\phi} \sinh\theta & \cosh \theta 
	\end{pmatrix}
	\begin{pmatrix}
	b \\ b^\dagger
	\end{pmatrix} \ .
\eeq
Then, by taking $\theta$ and $\phi$ such that
\begin{align}
	A &= 2D \cosh 2\theta \\
	B &= 2D  e^{-i\phi} \sinh 2\theta \ ,
\end{align}
where $D = \frac12 \sqrt{A^2 - |B|^2 }$, we have that 
\beq
\label{eq:rho3}
	\rho = C \exp\left[ -\frac{1}{2}\left(  A\cosh 2\theta - |B| \sinh 2\theta \right)(bb^\dagger + b^\dagger b) \right]
	= C \exp \left[ -D (bb^\dagger + b^\dagger b)  \right]
\eeq
In this form, we can calculate expectation values of various operators. We start with
\beq
	\langle a^2 \rangle = \Tr (a^2 \rho ) = \left\langle \cosh^2 \theta\ b^2 - \frac12 e^{i\phi} \sinh 2\theta (b^\dagger b + b b^\dagger) +
		e^{2i\phi} \sinh^2 \theta\ b^{\dagger 2} \right\rangle \ .
\eeq
First, we notice that
\beq
	\langle b^2 \rangle = \langle b^{\dagger 2} \rangle = 0,
\eeq
which one can see by inserting $b^2$ or $b^{\dagger 2}$ into the trace, expanding in a basis of 
eigenvectors of the number operator $N_b = b^\dagger b$, and using orthogonality.
Next,
from Eq.~\eqref{eq:rho3} we have 
\beq 
	\rho = C e^{-D (bb^\dagger + b^\dagger b)} = C e^{-D (2 b^\dagger b + 1)} =  C e^{-D} e^{-2 D  N_b } \ .
\eeq
First, we have that $\langle 1 \rangle = \Tr \rho = (C/2) (\sinh D)^{-1}$, which 
means that to normalize we should take $C = 2 \sinh D$.  
Next, we can compute
\begin{align}
	\langle b^\dagger b \rangle = \Tr (b^\dagger b \rho) = \Tr (b \rho b^\dagger) 
	&=  C e^{-D} \sum_{n= 0}^\infty \langle n | b e^{-2 D  N_b }  b^\dagger |n\rangle \\
	&=  C e^{-D} \sum_{n= 0}^\infty (n+1)  \langle n+1 |  e^{-2 D  N_b }  |n+1\rangle \\
	&=  C e^{-D} \sum_{n= 0}^\infty (n+1)  e^{-2 D  (n+1)} \\
	&=  C e^{-D} \left( e^{-2D}  \frac{ e^{-2D} }{(1- e^{-2 D})^2} +  e^{-2D}  \frac{1}{1- e^{-2D} } \right) \\
	&=  \frac{C}{4}  \frac{e^{-D}}{\sinh^2 D} \\
	&=  \frac{1}{2}  \frac{e^{-D}}{\sinh D} \ .
\end{align}
So we have
\beq
	\langle b^\dagger b + b b^\dagger \rangle = 2 \langle b^\dagger b \rangle + \langle 1 \rangle =
		 \frac{e^{-D}}{\sinh D} + 1 = \coth D.
\eeq
From here we can match to Eqs.~\eqref{eq:ev1} - \eqref{eq:ev3}: 
\beq
\label{eq:B1}
	\langle a^2 \rangle = -\frac{1}{2} e^{i \phi} \sinh 2\theta \coth D = -\frac{\bar B}{4D} \coth D = x
\eeq
and
\beq
\label{eq:A1}
	\langle a a^\dagger \rangle = \cosh^2 \theta \langle b b^\dagger \rangle + \sinh^2 \theta \langle b^\dagger b \rangle 
	= \frac{A}{4D} \coth D + \frac12 = y\ .
\eeq

Returning to the computation of the entropy, we have that 
\begin{align}
\label{eq:S1}
	S(\rho) &= -\langle \log \rho \rangle = -\langle \log C \rangle + \frac{A}{2} \langle aa^\dagger + a^\dagger a \rangle
	+ \frac{B}{2} \langle a^2 \rangle + \frac{\bar B}{2} \langle a^{\dagger 2} \rangle  \ .
\end{align}
We ultimately want $S(\rho)$ in terms of the original parameters $B_i$, which we will do by 
first writing things in terms of  
\beq
	\chi = \frac12 \coth D = \sqrt{(y-\frac12)^2 - |x|^2},
\eeq
from which one can then use Eq.~\eqref{eq:chi1}.
First, using hyperbolic trigonometric identities, one can show that 
$-\log C = -\log (2 \sinh D ) = \frac12 \log(\chi + \frac12) + \frac12 \log(\chi - \frac12)$.
Then, solve the final equations in Eqs.~\eqref{eq:B1} and \eqref{eq:A1} for $B$ and $A$, respectively, and 
substitute them into Eq.~\eqref{eq:S1}, along with the expressions for the expectation values 
in terms of $x$ and $y$. One ends up with 
\begin{align}
	S(\rho) &= \tfrac12 \log(\chi + \tfrac12) + \tfrac12 \log(\chi - \tfrac12) + \frac{4D}{\coth D} 
	\left( (y-\frac12)^2 - |x|^2 \right) \\
	&=  \tfrac12 \log(\chi + \tfrac12) + \tfrac12 \log(\chi - \tfrac12)  + D\coth D \\
	&=  \tfrac12 \log(\chi + \tfrac12) + \tfrac12 \log(\chi - \tfrac12)  + \chi \log\left( 
	\frac{\chi + \frac12}{\chi -\frac12} \right) \\
	&=  (\chi+ \tfrac12)\log (\chi+ \tfrac12) - (\chi- \tfrac12)\log (\chi- \tfrac12) \ ,
\end{align}
where the penultimate equality also comes from hyperbolic trigonometric identities.
This gives Eq.~\eqref{eq:vNEnt}.




\begin{thebibliography}{10}%
\makeatletter
\providecommand \@ifxundefined [1]{%
 \ifx #1\undefined \expandafter \@firstoftwo
 \else \expandafter \@secondoftwo
\fi
}%
\providecommand \@ifnum [1]{%
 \ifnum #1\expandafter \@firstoftwo
 \else \expandafter \@secondoftwo
\fi
}%
\providecommand \enquote [1]{``#1''}%
\providecommand \bibnamefont  [1]{#1}%
\providecommand \bibfnamefont [1]{#1}%
\providecommand \citenamefont [1]{#1}%
\providecommand\href[0]{\@sanitize\@href}%
\providecommand\@href[1]{\endgroup\@@startlink{#1}\endgroup\@@href}%
\providecommand\@@href[1]{#1\@@endlink}%
\providecommand \@sanitize [0]{\begingroup\catcode`\&12\catcode`\#12\relax}%
\@ifxundefined \pdfoutput {\@firstoftwo}{%
 \@ifnum{\z@=\pdfoutput}{\@firstoftwo}{\@secondoftwo}%
}{%
 \providecommand\@@startlink[1]{\leavevmode}%
 \providecommand\@@endlink[0]{}%
}{%
 \providecommand\@@startlink[1]{%
  \leavevmode
  \pdfstartlink
   attr{/Border[0 0 1 ]/H/I/C[0 1 1]}%
   user{/Subtype/Link/A<</Type/Action/S/URI/URI(#1)>>}%
  \relax
 }%
 \providecommand\@@endlink[0]{\pdfendlink}%
}%
\providecommand \url  [0]{\begingroup\@sanitize \@url }%
\providecommand \@url [1]{\endgroup\@href {#1}{\urlprefix}}%
\providecommand \urlprefix [0]{URL }%
\providecommand \Eprint[0]{\href }%
\@ifxundefined \urlstyle {%
  \providecommand \doi [1]{doi:\discretionary{}{}{}#1}%
}{%
  \providecommand \doi [0]{doi:\discretionary{}{}{}\begingroup
  \urlstyle{rm}\Url }%
}%
\providecommand \doibase [0]{http://dx.doi.org/}%
\providecommand \Doi[1]{\href{\doibase#1}}%
\providecommand \bibAnnote [3]{%
  \BibitemShut{#1}%
  \begin{quotation}\noindent
    \textsc{Key:}\ #2\\\textsc{Annotation:}\ #3%
  \end{quotation}%
}%
\providecommand \bibAnnoteFile [2]{%
  \IfFileExists{#2}{\bibAnnote {#1} {#2} {\input{#2}}}{}%
}%
\providecommand \typeout [0]{\immediate \write \m@ne }%
\providecommand \selectlanguage [0]{\@gobble}%
\providecommand \bibinfo [0]{\@secondoftwo}%
\providecommand \bibfield [0]{\@secondoftwo}%
\providecommand \translation [1]{[#1]}%
\providecommand \BibitemOpen[0]{}%
\providecommand \bibitemStop [0]{}%
\providecommand \bibitemNoStop [0]{.\EOS\space}%
\providecommand \EOS [0]{\spacefactor3000\relax}%
\providecommand \BibitemShut [1]{\csname bibitem#1\endcsname}%
\bibitem{Page:1993df}%
  \BibitemOpen
  \bibfield{author}{%
  \bibinfo {author} {\bibfnamefont{Don~N.}\ \bibnamefont{Page}},\ }%
  \bibfield{title}{%
  \enquote{\bibinfo {title} {{Average entropy of a subsystem}},}\ }%
  \bibfield{journal}{%
  \Doi{10.1103/PhysRevLett.71.1291}{\bibinfo {journal} {Phys.Rev.Lett.}}\ }%
  \textbf{\bibinfo {volume} {71}},\ \bibinfo {pages} {1291--1294} (\bibinfo
  {year} {1993}),\
  \Eprint{http://arxiv.org/abs/gr-qc/9305007}{arXiv:gr-qc/9305007 [gr-qc]}%
  \bibAnnoteFile{NoStop}{Page:1993df}%
\bibitem{2007PhRvA..76e2319B}%
  \BibitemOpen
  \bibfield{author}{%
  \bibinfo {author} {\bibfnamefont{S.}~\bibnamefont{{Bravyi}}},\ }%
  \bibfield{title}{%
  \enquote{\bibinfo {title} {{Upper bounds on entangling rates of bipartite
  Hamiltonians}},}\ }%
  \bibfield{journal}{%
  \Doi{10.1103/PhysRevA.76.052319}{\bibinfo {journal} {\pra}}\ }%
  \textbf{\bibinfo {volume} {76}},\ \bibinfo {eid} {052319} (\bibinfo {month}
  {Nov.}\ \bibinfo {year} {2007}),\
  \Eprint{http://arxiv.org/abs/0704.0964}{arXiv:0704.0964 [quant-ph]}%
  \bibAnnoteFile{NoStop}{2007PhRvA..76e2319B}%
\bibitem{2013PhRvL.111q0501V}%
  \BibitemOpen
  \bibfield{author}{%
  \bibinfo {author} {\bibfnamefont{K.}~\bibnamefont{{Van Acoleyen}}}, \bibinfo
  {author} {\bibfnamefont{M.}~\bibnamefont{{Mari{\"e}n}}},\ and\ \bibinfo
  {author} {\bibfnamefont{F.}~\bibnamefont{{Verstraete}}},\ }%
  \bibfield{title}{%
  \enquote{\bibinfo {title} {{Entanglement Rates and Area Laws}},}\ }%
  \bibfield{journal}{%
  \Doi{10.1103/PhysRevLett.111.170501}{\bibinfo {journal} {Physical Review
  Letters}}\ }%
  \textbf{\bibinfo {volume} {111}},\ \bibinfo {eid} {170501} (\bibinfo {month}
  {Oct.}\ \bibinfo {year} {2013}),\
  \Eprint{http://arxiv.org/abs/1304.5931}{arXiv:1304.5931 [quant-ph]}%
  \bibAnnoteFile{NoStop}{2013PhRvL.111q0501V}%
\bibitem{Avery:2014dba}%
  \BibitemOpen
  \bibfield{author}{%
  \bibinfo {author} {\bibfnamefont{Steven~G.}\ \bibnamefont{Avery}}\ and\
  \bibinfo {author} {\bibfnamefont{Miguel~F.}\ \bibnamefont{Paulos}},\ }%
  \bibfield{title}{%
  \enquote{\bibinfo {title} {{Universal Bounds on the Time Evolution of
  Entanglement Entropy}},}\ }%
  \bibfield{journal}{%
  \Doi{10.1103/PhysRevLett.113.231604}{\bibinfo {journal} {Phys.Rev.Lett.}}\ }%
  \textbf{\bibinfo {volume} {113}},\ \bibinfo {pages} {231604} (\bibinfo {year}
  {2014}),\ \Eprint{http://arxiv.org/abs/1407.0705}{arXiv:1407.0705 [hep-th]}%
  \bibAnnoteFile{NoStop}{Avery:2014dba}%
\bibitem{Bousso:2014uxa}%
  \BibitemOpen
  \bibfield{author}{%
  \bibinfo {author} {\bibfnamefont{Raphael}\ \bibnamefont{Bousso}}, \bibinfo
  {author} {\bibfnamefont{Horacio}\ \bibnamefont{Casini}}, \bibinfo {author}
  {\bibfnamefont{Zachary}\ \bibnamefont{Fisher}},\ and\ \bibinfo {author}
  {\bibfnamefont{Juan}\ \bibnamefont{Maldacena}},\ }%
  \bibfield{title}{%
  \enquote{\bibinfo {title} {{Entropy on a null surface for interacting quantum
  field theories and the Bousso bound}},}\ }%
   (\bibinfo {year} {2014}),\
  \Eprint{http://arxiv.org/abs/1406.4545}{arXiv:1406.4545 [hep-th]}%
  \bibAnnoteFile{NoStop}{Bousso:2014uxa}%
\bibitem{Zurek:1994wd}%
  \BibitemOpen
  \bibfield{author}{%
  \bibinfo {author} {\bibfnamefont{Wojciech~Hubert}\ \bibnamefont{Zurek}}\ and\
  \bibinfo {author} {\bibfnamefont{Juan~Pablo}\ \bibnamefont{Paz}},\ }%
  \bibfield{title}{%
  \enquote{\bibinfo {title} {{Decoherence, chaos, and the second law}},}\ }%
  \bibfield{journal}{%
  \Doi{10.1103/PhysRevLett.72.2508}{\bibinfo {journal} {Phys.Rev.Lett.}}\ }%
  \textbf{\bibinfo {volume} {72}},\ \bibinfo {pages} {2508} (\bibinfo {year}
  {1994}),\ \Eprint{http://arxiv.org/abs/gr-qc/9402006}{arXiv:gr-qc/9402006
  [gr-qc]}%
  \bibAnnoteFile{NoStop}{Zurek:1994wd}%
\bibitem{Zurek:1995jd}%
  \BibitemOpen
  \bibfield{author}{%
  \bibinfo {author} {\bibfnamefont{Wojciech~Hubert}\ \bibnamefont{Zurek}}\ and\
  \bibinfo {author} {\bibfnamefont{Juan~Pablo}\ \bibnamefont{Paz}},\ }%
  \bibfield{title}{%
  \enquote{\bibinfo {title} {{Quantum chaos: A decoherent definition}},}\ }%
  \bibfield{journal}{%
  \Doi{10.1016/0167-2789(94)00271-Q}{\bibinfo {journal} {Physica}}\ }%
  \textbf{\bibinfo {volume} {D83}},\ \bibinfo {pages} {300} (\bibinfo {year}
  {1995}),\
  \Eprint{http://arxiv.org/abs/quant-ph/9502029}{arXiv:quant-ph/9502029
  [quant-ph]}%
  \bibAnnoteFile{NoStop}{Zurek:1995jd}%
\bibitem{PhysRevLett.83.4526}%
  \BibitemOpen
  \bibfield{author}{%
  \bibinfo {author} {\bibfnamefont{Arjendu~K.}\ \bibnamefont{Pattanayak}},\ }%
  \bibfield{title}{%
  \enquote{\bibinfo {title} {Lyapunov exponents, entropy production, and
  decoherence},}\ }%
  \bibfield{journal}{%
  \Doi{10.1103/PhysRevLett.83.4526}{\bibinfo {journal} {Phys. Rev. Lett.}}\ }%
  \textbf{\bibinfo {volume} {83}},\ \bibinfo {pages} {4526--4529} (\bibinfo
  {month} {Nov}\ \bibinfo {year} {1999}),\
  \Eprint{http://arxiv.org/abs/arXiv:chao-dyn/9911017}{arXiv:chao-dyn/9911017}%
  \bibAnnoteFile{NoStop}{PhysRevLett.83.4526}%
\bibitem{PhysRevE.60.1542}%
  \BibitemOpen
  \bibfield{author}{%
  \bibinfo {author} {\bibfnamefont{Paul~A.}\ \bibnamefont{Miller}}\ and\
  \bibinfo {author} {\bibfnamefont{Sarben}\ \bibnamefont{Sarkar}},\ }%
  \bibfield{title}{%
  \enquote{\bibinfo {title} {Signatures of chaos in the entanglement of two
  coupled quantum kicked tops},}\ }%
  \bibfield{journal}{%
  \Doi{10.1103/PhysRevE.60.1542}{\bibinfo {journal} {Phys. Rev. E}}\ }%
  \textbf{\bibinfo {volume} {60}},\ \bibinfo {pages} {1542--1550} (\bibinfo
  {month} {Aug}\ \bibinfo {year} {1999})%
  \bibAnnoteFile{NoStop}{PhysRevE.60.1542}%
\bibitem{2000PhRvL..85.3373M}%
  \BibitemOpen
  \bibfield{author}{%
  \bibinfo {author} {\bibfnamefont{D.}~\bibnamefont{{Monteoliva}}}\ and\
  \bibinfo {author} {\bibfnamefont{J.~P.}\ \bibnamefont{{Paz}}},\ }%
  \bibfield{title}{%
  \enquote{\bibinfo {title} {{Decoherence and the Rate of Entropy Production in
  Chaotic Quantum Systems}},}\ }%
  \bibfield{journal}{%
  \Doi{10.1103/PhysRevLett.85.3373}{\bibinfo {journal} {Physical Review
  Letters}}\ }%
  \textbf{\bibinfo {volume} {85}},\ \bibinfo {pages} {3373} (\bibinfo {month}
  {Oct.}\ \bibinfo {year} {2000}),\
  \Eprint{http://arxiv.org/abs/quant-ph/0007052}{quant-ph/0007052}%
  \bibAnnoteFile{NoStop}{2000PhRvL..85.3373M}%
\bibitem{2002PhRvE..66d5201T}%
  \BibitemOpen
  \bibfield{author}{%
  \bibinfo {author} {\bibfnamefont{A.}~\bibnamefont{{Tanaka}}}, \bibinfo
  {author} {\bibfnamefont{H.}~\bibnamefont{{Fujisaki}}},\ and\ \bibinfo
  {author} {\bibfnamefont{T.}~\bibnamefont{{Miyadera}}},\ }%
  \bibfield{title}{%
  \enquote{\bibinfo {title} {{Saturation of the production of quantum
  entanglement between weakly coupled mapping systems in a strongly chaotic
  region}},}\ }%
  \bibfield{journal}{%
  \Doi{10.1103/PhysRevE.66.045201}{\bibinfo {journal} {\pre}}\ }%
  \textbf{\bibinfo {volume} {66}},\ \bibinfo {eid} {045201} (\bibinfo {month}
  {Oct.}\ \bibinfo {year} {2002}),\
  \Eprint{http://arxiv.org/abs/quant-ph/0209086}{quant-ph/0209086}%
  \bibAnnoteFile{NoStop}{2002PhRvE..66d5201T}%
\bibitem{2003PhRvE..67f6201F}%
  \BibitemOpen
  \bibfield{author}{%
  \bibinfo {author} {\bibfnamefont{H.}~\bibnamefont{{Fujisaki}}}, \bibinfo
  {author} {\bibfnamefont{T.}~\bibnamefont{{Miyadera}}},\ and\ \bibinfo
  {author} {\bibfnamefont{A.}~\bibnamefont{{Tanaka}}},\ }%
  \bibfield{title}{%
  \enquote{\bibinfo {title} {{Dynamical aspects of quantum entanglement for
  weakly coupled kicked tops}},}\ }%
  \bibfield{journal}{%
  \Doi{10.1103/PhysRevE.67.066201}{\bibinfo {journal} {\pre}}\ }%
  \textbf{\bibinfo {volume} {67}},\ \bibinfo {eid} {066201} (\bibinfo {month}
  {Jun.}\ \bibinfo {year} {2003}),\
  \Eprint{http://arxiv.org/abs/quant-ph/0211110}{quant-ph/0211110}%
  \bibAnnoteFile{NoStop}{2003PhRvE..67f6201F}%
\bibitem{2003JPSJ...72S.111F}%
  \BibitemOpen
  \bibfield{author}{%
  \bibinfo {author} {\bibfnamefont{H.}~\bibnamefont{{Fujisaki}}}, \bibinfo
  {author} {\bibfnamefont{A.}~\bibnamefont{{Tanaka}}},\ and\ \bibinfo {author}
  {\bibfnamefont{T.}~\bibnamefont{{Miyadera}}},\ }%
  \bibfield{title}{%
  \enquote{\bibinfo {title} {{Dynamical Aspects of Quantum Entanglement for
  Coupled Mapping Systems}},}\ }%
  \bibfield{journal}{%
  \Doi{10.1143/JPSJS.72SC.111}{\bibinfo {journal} {Journal of the Physical
  Society of Japan}}\ }%
  \textbf{\bibinfo {volume} {72}},\ \bibinfo {pages} {111--114} (\bibinfo
  {month} {Jan.}\ \bibinfo {year} {2003}),\
  \Eprint{http://arxiv.org/abs/quant-ph/0302015}{quant-ph/0302015}%
  \bibAnnoteFile{NoStop}{2003JPSJ...72S.111F}%
\bibitem{2003PhRvA..68c2104B}%
  \BibitemOpen
  \bibfield{author}{%
  \bibinfo {author} {\bibfnamefont{R.}~\bibnamefont{{Blume-Kohout}}}\ and\
  \bibinfo {author} {\bibfnamefont{W.~H.}\ \bibnamefont{{Zurek}}},\ }%
  \bibfield{title}{%
  \enquote{\bibinfo {title} {{Decoherence from a chaotic environment: An
  upside-down ``oscillator'' as a model}},}\ }%
  \bibfield{journal}{%
  \Doi{10.1103/PhysRevA.68.032104}{\bibinfo {journal} {Phys. Rev. A}}\ }%
  \textbf{\bibinfo {volume} {68}},\ \bibinfo {eid} {032104} (\bibinfo {month}
  {Sep.}\ \bibinfo {year} {2003}),\
  \Eprint{http://arxiv.org/abs/arXiv:quant-ph/0212153}{arXiv:quant-ph/0212153}%
  \bibAnnoteFile{NoStop}{2003PhRvA..68c2104B}%
\bibitem{2003JPhA...36.2463Z}%
  \BibitemOpen
  \bibfield{author}{%
  \bibinfo {author} {\bibfnamefont{M.}~\bibnamefont{{Znidaric}}}\ and\ \bibinfo
  {author} {\bibfnamefont{T.}~\bibnamefont{{Prosen}}},\ }%
  \bibfield{title}{%
  \enquote{\bibinfo {title} {{Fidelity and purity decay in weakly coupled
  composite systems}},}\ }%
  \bibfield{journal}{%
  \Doi{10.1088/0305-4470/36/10/307}{\bibinfo {journal} {Journal of Physics A
  Mathematical General}}\ }%
  \textbf{\bibinfo {volume} {36}},\ \bibinfo {pages} {2463--2481} (\bibinfo
  {month} {Mar.}\ \bibinfo {year} {2003}),\
  \Eprint{http://arxiv.org/abs/quant-ph/0209145}{quant-ph/0209145}%
  \bibAnnoteFile{NoStop}{2003JPhA...36.2463Z}%
\bibitem{2004JPhA...37.5157A}%
  \BibitemOpen
  \bibfield{author}{%
  \bibinfo {author} {\bibfnamefont{R.}~\bibnamefont{{Alicki}}}, \bibinfo
  {author} {\bibfnamefont{A.}~\bibnamefont{{Lozinski}}}, \bibinfo {author}
  {\bibfnamefont{P.}~\bibnamefont{{Pakonski}}},\ and\ \bibinfo {author}
  {\bibfnamefont{K.}~\bibnamefont{{Zyczkowski}}},\ }%
  \bibfield{title}{%
  \enquote{\bibinfo {title} {{Quantum dynamical entropy and decoherence
  rate}},}\ }%
  \bibfield{journal}{%
  \Doi{10.1088/0305-4470/37/19/004}{\bibinfo {journal} {Journal of Physics A
  Mathematical General}}\ }%
  \textbf{\bibinfo {volume} {37}},\ \bibinfo {pages} {5157--5172} (\bibinfo
  {month} {May}\ \bibinfo {year} {2004}),\
  \Eprint{http://arxiv.org/abs/arXiv:quant-ph/0309194}{arXiv:quant-ph/0309194}%
  \bibAnnoteFile{NoStop}{2004JPhA...37.5157A}%
\bibitem{2004PhRvL..92o0403J}%
  \BibitemOpen
  \bibfield{author}{%
  \bibinfo {author} {\bibfnamefont{P.}~\bibnamefont{{Jacquod}}},\ }%
  \bibfield{title}{%
  \enquote{\bibinfo {title} {{Semiclassical Time Evolution of the Reduced
  Density Matrix and Dynamically Assisted Generation of Entanglement for
  Bipartite Quantum Systems}},}\ }%
  \bibfield{journal}{%
  \Doi{10.1103/PhysRevLett.92.150403}{\bibinfo {journal} {Physical Review
  Letters}}\ }%
  \textbf{\bibinfo {volume} {92}},\ \bibinfo {eid} {150403} (\bibinfo {month}
  {Apr.}\ \bibinfo {year} {2004}),\
  \Eprint{http://arxiv.org/abs/quant-ph/0308099}{quant-ph/0308099}%
  \bibAnnoteFile{NoStop}{2004PhRvL..92o0403J}%
\bibitem{PhysRevLett.93.219903}%
  \BibitemOpen
  \bibfield{author}{%
  \bibinfo {author} {\bibfnamefont{Ph.}\ \bibnamefont{Jacquod}},\ }%
  \bibfield{title}{%
  \enquote{\bibinfo {title} {Erratum: Semiclassical time evolution of the
  reduced density matrix and dynamically assisted generation of entanglement
  for bipartite quantum systems [phys. rev. lett. \textbf{92} , 150403
  (2004)]},}\ }%
  \bibfield{journal}{%
  \Doi{10.1103/PhysRevLett.93.219903}{\bibinfo {journal} {Phys. Rev. Lett.}}\
  }%
  \textbf{\bibinfo {volume} {93}},\ \bibinfo {pages} {219903} (\bibinfo {month}
  {Nov}\ \bibinfo {year} {2004})%
  \bibAnnoteFile{NoStop}{PhysRevLett.93.219903}%
\bibitem{2006PhRvL..97s4103P}%
  \BibitemOpen
  \bibfield{author}{%
  \bibinfo {author} {\bibfnamefont{C.}~\bibnamefont{{Petitjean}}}\ and\
  \bibinfo {author} {\bibfnamefont{P.}~\bibnamefont{{Jacquod}}},\ }%
  \bibfield{title}{%
  \enquote{\bibinfo {title} {{Lyapunov Generation of Entanglement and the
  Correspondence Principle}},}\ }%
  \bibfield{journal}{%
  \Doi{10.1103/PhysRevLett.97.194103}{\bibinfo {journal} {Physical Review
  Letters}}\ }%
  \textbf{\bibinfo {volume} {97}},\ \bibinfo {eid} {194103} (\bibinfo {month}
  {Nov.}\ \bibinfo {year} {2006}),\
  \Eprint{http://arxiv.org/abs/quant-ph/0510157}{quant-ph/0510157}%
  \bibAnnoteFile{NoStop}{2006PhRvL..97s4103P}%
\bibitem{2009AdPhy..58...67J}%
  \BibitemOpen
  \bibfield{author}{%
  \bibinfo {author} {\bibfnamefont{P.}~\bibnamefont{{Jacquod}}}\ and\ \bibinfo
  {author} {\bibfnamefont{C.}~\bibnamefont{{Petitjean}}},\ }%
  \bibfield{title}{%
  \enquote{\bibinfo {title} {{Decoherence, entanglement and irreversibility in
  quantum dynamical systems with few degrees of freedom}},}\ }%
  \bibfield{journal}{%
  \Doi{10.1080/00018730902831009}{\bibinfo {journal} {Advances in Physics}}\ }%
  \textbf{\bibinfo {volume} {58}},\ \bibinfo {pages} {67--196} (\bibinfo
  {month} {Mar.}\ \bibinfo {year} {2009}),\
  \Eprint{http://arxiv.org/abs/0806.0987}{arXiv:0806.0987 [quant-ph]}%
  \bibAnnoteFile{NoStop}{2009AdPhy..58...67J}%
\bibitem{2008JPhA...41k5303F}%
  \BibitemOpen
  \bibfield{author}{%
  \bibinfo {author} {\bibfnamefont{K.~M.}\ \bibnamefont{{Fonseca Romero}}},
  \bibinfo {author} {\bibfnamefont{J.~E.}\ \bibnamefont{{Parreira}}}, \bibinfo
  {author} {\bibfnamefont{L.~A.~M.}\ \bibnamefont{{Souza}}}, \bibinfo {author}
  {\bibfnamefont{M.~C.}\ \bibnamefont{{Nemes}}},\ and\ \bibinfo {author}
  {\bibfnamefont{W.}~\bibnamefont{{Wreszinski}}},\ }%
  \bibfield{title}{%
  \enquote{\bibinfo {title} {{An analytical relation between entropy production
  and quantum Lyapunov exponents for Gaussian bipartite systems}},}\ }%
  \bibfield{journal}{%
  \Doi{10.1088/1751-8113/41/11/115303}{\bibinfo {journal} {Journal of Physics A
  Mathematical General}}\ }%
  \textbf{\bibinfo {volume} {41}},\ \bibinfo {pages} {115303} (\bibinfo {month}
  {Mar.}\ \bibinfo {year} {2008}),\
  \Eprint{http://arxiv.org/abs/arXiv:quant-ph/0703200}{arXiv:quant-ph/0703200}%
  \bibAnnoteFile{NoStop}{2008JPhA...41k5303F}%
\bibitem{2014OptCo.331..148S}%
  \BibitemOpen
  \bibfield{author}{%
  \bibinfo {author} {\bibfnamefont{L.~A.~M.}\ \bibnamefont{{Souza}}}, \bibinfo
  {author} {\bibfnamefont{J.~G.~P.}\ \bibnamefont{{Faria}}},\ and\ \bibinfo
  {author} {\bibfnamefont{M.~C.}\ \bibnamefont{{Nemes}}},\ }%
  \bibfield{title}{%
  \enquote{\bibinfo {title} {{Parametric competition in non-autonomous
  Hamiltonian systems}},}\ }%
  \bibfield{journal}{%
  \Doi{10.1016/j.optcom.2014.05.070}{\bibinfo {journal} {Optics
  Communications}}\ }%
  \textbf{\bibinfo {volume} {331}},\ \bibinfo {pages} {148--153} (\bibinfo
  {month} {Nov.}\ \bibinfo {year} {2014}),\
  \Eprint{http://arxiv.org/abs/1405.7955}{arXiv:1405.7955 [quant-ph]}%
  \bibAnnoteFile{NoStop}{2014OptCo.331..148S}%
\bibitem{2010PhRvA..82e2335R}%
  \BibitemOpen
  \bibfield{author}{%
  \bibinfo {author} {\bibfnamefont{A.~D.}\ \bibnamefont{{Ribeiro}}}\ and\
  \bibinfo {author} {\bibfnamefont{R.~M.}\ \bibnamefont{{Angelo}}},\ }%
  \bibfield{title}{%
  \enquote{\bibinfo {title} {{Entanglement dynamics via coherent-state
  propagators}},}\ }%
  \bibfield{journal}{%
  \Doi{10.1103/PhysRevA.82.052335}{\bibinfo {journal} {\pra}}\ }%
  \textbf{\bibinfo {volume} {82}},\ \bibinfo {eid} {052335} (\bibinfo {month}
  {Nov.}\ \bibinfo {year} {2010}),\
  \Eprint{http://arxiv.org/abs/1009.2079}{arXiv:1009.2079 [quant-ph]}%
  \bibAnnoteFile{NoStop}{2010PhRvA..82e2335R}%
\bibitem{2011PhRvE..83d6214B}%
  \BibitemOpen
  \bibfield{author}{%
  \bibinfo {author} {\bibfnamefont{M.~V.~S.}\ \bibnamefont{{Bonan{\c c}a}}},\
  }%
  \bibfield{title}{%
  \enquote{\bibinfo {title} {{Lyapunov decoherence rate in classically chaotic
  systems}},}\ }%
  \bibfield{journal}{%
  \Doi{10.1103/PhysRevE.83.046214}{\bibinfo {journal} {\pre}}\ }%
  \textbf{\bibinfo {volume} {83}},\ \bibinfo {eid} {046214} (\bibinfo {month}
  {Apr.}\ \bibinfo {year} {2011}),\
  \Eprint{http://arxiv.org/abs/1010.2115}{arXiv:1010.2115 [quant-ph]}%
  \bibAnnoteFile{NoStop}{2011PhRvE..83d6214B}%
\bibitem{Calabrese:2005in}%
  \BibitemOpen
  \bibfield{author}{%
  \bibinfo {author} {\bibfnamefont{Pasquale}\ \bibnamefont{Calabrese}}\ and\
  \bibinfo {author} {\bibfnamefont{John~L.}\ \bibnamefont{Cardy}},\ }%
  \bibfield{title}{%
  \enquote{\bibinfo {title} {{Evolution of entanglement entropy in
  one-dimensional systems}},}\ }%
  \bibfield{journal}{%
  \Doi{10.1088/1742-5468/2005/04/P04010}{\bibinfo {journal} {J.Stat.Mech.}}\ }%
  \textbf{\bibinfo {volume} {0504}},\ \bibinfo {pages} {P04010} (\bibinfo
  {year} {2005}),\
  \Eprint{http://arxiv.org/abs/cond-mat/0503393}{arXiv:cond-mat/0503393
  [cond-mat]}%
  \bibAnnoteFile{NoStop}{Calabrese:2005in}%
\bibitem{Calabrese:2007rg}%
  \BibitemOpen
  \bibfield{author}{%
  \bibinfo {author} {\bibfnamefont{Pasquale}\ \bibnamefont{Calabrese}}\ and\
  \bibinfo {author} {\bibfnamefont{John}\ \bibnamefont{Cardy}},\ }%
  \bibfield{title}{%
  \enquote{\bibinfo {title} {{Quantum Quenches in Extended Systems}},}\ }%
  \bibfield{journal}{%
  \Doi{10.1088/1742-5468/2007/06/P06008}{\bibinfo {journal} {J.Stat.Mech.}}\ }%
  \textbf{\bibinfo {volume} {0706}},\ \bibinfo {pages} {P06008} (\bibinfo
  {year} {2007}),\ \Eprint{http://arxiv.org/abs/0704.1880}{arXiv:0704.1880
  [cond-mat.stat-mech]}%
  \bibAnnoteFile{NoStop}{Calabrese:2007rg}%
\bibitem{Hubeny:2007xt}%
  \BibitemOpen
  \bibfield{author}{%
  \bibinfo {author} {\bibfnamefont{Veronika~E.}\ \bibnamefont{Hubeny}},
  \bibinfo {author} {\bibfnamefont{Mukund}\ \bibnamefont{Rangamani}},\ and\
  \bibinfo {author} {\bibfnamefont{Tadashi}\ \bibnamefont{Takayanagi}},\ }%
  \bibfield{title}{%
  \enquote{\bibinfo {title} {{A Covariant holographic entanglement entropy
  proposal}},}\ }%
  \bibfield{journal}{%
  \Doi{10.1088/1126-6708/2007/07/062}{\bibinfo {journal} {JHEP}}\ }%
  \textbf{\bibinfo {volume} {0707}},\ \bibinfo {pages} {062} (\bibinfo {year}
  {2007}),\ \Eprint{http://arxiv.org/abs/0705.0016}{arXiv:0705.0016 [hep-th]}%
  \bibAnnoteFile{NoStop}{Hubeny:2007xt}%
\bibitem{AbajoArrastia:2010yt}%
  \BibitemOpen
  \bibfield{author}{%
  \bibinfo {author} {\bibfnamefont{Javier}\ \bibnamefont{Abajo-Arrastia}},
  \bibinfo {author} {\bibfnamefont{Jo\~{a}o}\ \bibnamefont{Apar\'{i}cio}},\
  and\ \bibinfo {author} {\bibfnamefont{Esperanza}\ \bibnamefont{L\'{o}pez}},\
  }%
  \bibfield{title}{%
  \enquote{\bibinfo {title} {{Holographic Evolution of Entanglement
  Entropy}},}\ }%
  \bibfield{journal}{%
  \Doi{10.1007/JHEP11(2010)149}{\bibinfo {journal} {JHEP}}\ }%
  \textbf{\bibinfo {volume} {1011}},\ \bibinfo {pages} {149} (\bibinfo {year}
  {2010}),\ \Eprint{http://arxiv.org/abs/1006.4090}{arXiv:1006.4090 [hep-th]}%
  \bibAnnoteFile{NoStop}{AbajoArrastia:2010yt}%
\bibitem{Maldacena:1997re}%
  \BibitemOpen
  \bibfield{author}{%
  \bibinfo {author} {\bibfnamefont{Juan~Martin}\ \bibnamefont{Maldacena}},\ }%
  \bibfield{title}{%
  \enquote{\bibinfo {title} {{The Large N limit of superconformal field
  theories and supergravity}},}\ }%
  \bibfield{journal}{%
  \Doi{10.1023/A:1026654312961}{\bibinfo {journal} {Adv.Theor.Math.Phys.}}\ }%
  \textbf{\bibinfo {volume} {2}},\ \bibinfo {pages} {231--252} (\bibinfo {year}
  {1998}),\ \Eprint{http://arxiv.org/abs/hep-th/9711200}{arXiv:hep-th/9711200
  [hep-th]}%
  \bibAnnoteFile{NoStop}{Maldacena:1997re}%
\bibitem{Hayden:2007cs}%
  \BibitemOpen
  \bibfield{author}{%
  \bibinfo {author} {\bibfnamefont{Patrick}\ \bibnamefont{Hayden}}\ and\
  \bibinfo {author} {\bibfnamefont{John}\ \bibnamefont{Preskill}},\ }%
  \bibfield{title}{%
  \enquote{\bibinfo {title} {{Black holes as mirrors: Quantum information in
  random subsystems}},}\ }%
  \bibfield{journal}{%
  \Doi{10.1088/1126-6708/2007/09/120}{\bibinfo {journal} {JHEP}}\ }%
  \textbf{\bibinfo {volume} {0709}},\ \bibinfo {pages} {120} (\bibinfo {year}
  {2007}),\ \Eprint{http://arxiv.org/abs/0708.4025}{arXiv:0708.4025 [hep-th]}%
  \bibAnnoteFile{NoStop}{Hayden:2007cs}%
\bibitem{Sekino:2008he}%
  \BibitemOpen
  \bibfield{author}{%
  \bibinfo {author} {\bibfnamefont{Yasuhiro}\ \bibnamefont{Sekino}}\ and\
  \bibinfo {author} {\bibfnamefont{Leonard}\ \bibnamefont{Susskind}},\ }%
  \bibfield{title}{%
  \enquote{\bibinfo {title} {{Fast Scramblers}},}\ }%
  \bibfield{journal}{%
  \Doi{10.1088/1126-6708/2008/10/065}{\bibinfo {journal} {JHEP}}\ }%
  \textbf{\bibinfo {volume} {0810}},\ \bibinfo {pages} {065} (\bibinfo {year}
  {2008}),\ \Eprint{http://arxiv.org/abs/0808.2096}{arXiv:0808.2096 [hep-th]}%
  \bibAnnoteFile{NoStop}{Sekino:2008he}%
\bibitem{Gutzwiller:1971fy}%
  \BibitemOpen
  \bibfield{author}{%
  \bibinfo {author} {\bibfnamefont{M.C.}\ \bibnamefont{Gutzwiller}},\ }%
  \bibfield{title}{%
  \enquote{\bibinfo {title} {{Periodic orbits and classical quantization
  conditions}},}\ }%
  \bibfield{journal}{%
  \Doi{10.1063/1.1665596}{\bibinfo {journal} {J.Math.Phys.}}\ }%
  \textbf{\bibinfo {volume} {12}},\ \bibinfo {pages} {343--358} (\bibinfo
  {year} {1971})%
  \bibAnnoteFile{NoStop}{Gutzwiller:1971fy}%
\bibitem{Gutzwiller1991}%
  \BibitemOpen
  \bibfield{author}{%
  \bibinfo {author} {\bibfnamefont{Martin}\ \bibnamefont{Gutzwiller}},\ }%
  \emph{\bibinfo {title} {{Chaos in Classical and Quantum Mechanics}}}\
  (\bibinfo {publisher} {Springer},\ \bibinfo {address} {New York},\ \bibinfo
  {year} {1991})%
  \bibAnnoteFile{NoStop}{Gutzwiller1991}%
\bibitem{Cvitanovic2012}%
  \BibitemOpen
  \bibfield{author}{%
  \bibinfo {author} {\bibfnamefont{P.}~\bibnamefont{Cvitanovi\'{c}}}, \bibinfo
  {author} {\bibfnamefont{R.}~\bibnamefont{Artuso}}, \bibinfo {author}
  {\bibfnamefont{R.}~\bibnamefont{Mainieri}}, \bibinfo {author}
  {\bibfnamefont{G.}~\bibnamefont{Tanner}},\ and\ \bibinfo {author}
  {\bibfnamefont{G.}~\bibnamefont{Vattay}},\ }%
  \emph{\bibinfo {title} {{Chaos: Classical and Quantum}}}\ (\bibinfo
  {publisher} {Niels Bohr Institute},\ \bibinfo {address} {Copenhagen},\
  \bibinfo {year} {2012})\ \url{http://ChaosBook.org}%
  \bibAnnoteFile{NoStop}{Cvitanovic2012}%
\bibitem{Zaslavsky2005}%
  \BibitemOpen
  \bibfield{author}{%
  \bibinfo {author} {\bibfnamefont{George~M.}\ \bibnamefont{Zaslavsky}},\ }%
  \emph{\bibinfo {title} {{Hamiltonian Chaos and Fractional Dynamics}}}\
  (\bibinfo {publisher} {Oxford University Press},\ \bibinfo {address}
  {Oxford},\ \bibinfo {year} {2005})%
  \bibAnnoteFile{NoStop}{Zaslavsky2005}%
\bibitem{Zaslavsky2007}%
  \BibitemOpen
  \bibfield{author}{%
  \bibinfo {author} {\bibfnamefont{George~M.}\ \bibnamefont{Zaslavsky}},\ }%
  \emph{\bibinfo {title} {{The Physics of Chaos in Hamiltonian Systems, 2nd
  Ed.}}}\ (\bibinfo {publisher} {Imperial College Press},\ \bibinfo {address}
  {London},\ \bibinfo {year} {2007})%
  \bibAnnoteFile{NoStop}{Zaslavsky2007}%
\bibitem{PhysRevLett.80.5524}%
  \BibitemOpen
  \bibfield{author}{%
  \bibinfo {author} {\bibfnamefont{K.}~\bibnamefont{Furuya}}, \bibinfo {author}
  {\bibfnamefont{M.~C.}\ \bibnamefont{Nemes}},\ and\ \bibinfo {author}
  {\bibfnamefont{G.~Q.}\ \bibnamefont{Pellegrino}},\ }%
  \bibfield{title}{%
  \enquote{\bibinfo {title} {Quantum dynamical manifestation of chaotic
  behavior in the process of entanglement},}\ }%
  \bibfield{journal}{%
  \Doi{10.1103/PhysRevLett.80.5524}{\bibinfo {journal} {Phys. Rev. Lett.}}\ }%
  \textbf{\bibinfo {volume} {80}},\ \bibinfo {pages} {5524--5527} (\bibinfo
  {month} {Jun}\ \bibinfo {year} {1998})%
  \bibAnnoteFile{NoStop}{PhysRevLett.80.5524}%
\bibitem{2012PhRvE..85c6208C}%
  \BibitemOpen
  \bibfield{author}{%
  \bibinfo {author} {\bibfnamefont{G.}~\bibnamefont{{Casati}}}, \bibinfo
  {author} {\bibfnamefont{I.}~\bibnamefont{{Guarneri}}},\ and\ \bibinfo
  {author} {\bibfnamefont{J.}~\bibnamefont{{Reslen}}},\ }%
  \bibfield{title}{%
  \enquote{\bibinfo {title} {{Classical dynamics of quantum entanglement}},}\
  }%
  \bibfield{journal}{%
  \Doi{10.1103/PhysRevE.85.036208}{\bibinfo {journal} {\pre}}\ }%
  \textbf{\bibinfo {volume} {85}},\ \bibinfo {eid} {036208} (\bibinfo {month}
  {Mar.}\ \bibinfo {year} {2012}),\
  \Eprint{http://arxiv.org/abs/1109.0907}{arXiv:1109.0907 [quant-ph]}%
  \bibAnnoteFile{NoStop}{2012PhRvE..85c6208C}%
\bibitem{2005PhRvA..71d2321A}%
  \BibitemOpen
  \bibfield{author}{%
  \bibinfo {author} {\bibfnamefont{R.~M.}\ \bibnamefont{{Angelo}}}\ and\
  \bibinfo {author} {\bibfnamefont{K.}~\bibnamefont{{Furuya}}},\ }%
  \bibfield{title}{%
  \enquote{\bibinfo {title} {{Semiclassical limit of the entanglement in closed
  pure systems}},}\ }%
  \bibfield{journal}{%
  \Doi{10.1103/PhysRevA.71.042321}{\bibinfo {journal} {\pra}}\ }%
  \textbf{\bibinfo {volume} {71}},\ \bibinfo {eid} {042321} (\bibinfo {month}
  {Apr.}\ \bibinfo {year} {2005}),\
  \Eprint{http://arxiv.org/abs/quant-ph/0410103}{quant-ph/0410103}%
  \bibAnnoteFile{NoStop}{2005PhRvA..71d2321A}%
\bibitem{Hawking:1974sw}%
  \BibitemOpen
  \bibfield{author}{%
  \bibinfo {author} {\bibfnamefont{S.W.}\ \bibnamefont{Hawking}},\ }%
  \bibfield{title}{%
  \enquote{\bibinfo {title} {{Particle Creation by Black Holes}},}\ }%
  \bibfield{journal}{%
  \Doi{10.1007/BF02345020, 10.1007/BF02345020}{\bibinfo {journal}
  {Commun.Math.Phys.}}\ }%
  \textbf{\bibinfo {volume} {43}},\ \bibinfo {pages} {199--220} (\bibinfo
  {year} {1975})%
  \bibAnnoteFile{NoStop}{Hawking:1974sw}%
\bibitem{2000JPhA...33.8139B}%
  \BibitemOpen
  \bibfield{author}{%
  \bibinfo {author} {\bibfnamefont{F.}~\bibnamefont{{Benatti}}}\ and\ \bibinfo
  {author} {\bibfnamefont{R.}~\bibnamefont{{Floreanini}}},\ }%
  \bibfield{title}{%
  \enquote{\bibinfo {title} {{Damped harmonic oscillators in the holomorphic
  representation}},}\ }%
  \bibfield{journal}{%
  \Doi{10.1088/0305-4470/33/45/310}{\bibinfo {journal} {Journal of Physics A
  Mathematical General}}\ }%
  \textbf{\bibinfo {volume} {33}},\ \bibinfo {pages} {8139--8153} (\bibinfo
  {month} {Nov.}\ \bibinfo {year} {2000}),\
  \Eprint{http://arxiv.org/abs/arXiv:hep-th/0010013}{arXiv:hep-th/0010013}%
  \bibAnnoteFile{NoStop}{2000JPhA...33.8139B}%
\bibitem{Jose1998}%
  \BibitemOpen
  \bibfield{author}{%
  \bibinfo {author} {\bibfnamefont{Jorge~V.}\ \bibnamefont{Jos\'e}}\ and\
  \bibinfo {author} {\bibfnamefont{Eugene~J.}\ \bibnamefont{Saletan}},\ }%
  \emph{\bibinfo {title} {{Classical Dynamics: A Contemporary Approach}}}\
  (\bibinfo {publisher} {Cambridge University Press},\ \bibinfo {address}
  {Cambridge},\ \bibinfo {year} {1998})%
  \bibAnnoteFile{NoStop}{Jose1998}%
\bibitem{2007PhRvL..99m0601G}%
  \BibitemOpen
  \bibfield{author}{%
  \bibinfo {author} {\bibfnamefont{F.}~\bibnamefont{{Ginelli}}}, \bibinfo
  {author} {\bibfnamefont{P.}~\bibnamefont{{Poggi}}}, \bibinfo {author}
  {\bibfnamefont{A.}~\bibnamefont{{Turchi}}}, \bibinfo {author}
  {\bibfnamefont{H.}~\bibnamefont{{Chat{\'e}}}}, \bibinfo {author}
  {\bibfnamefont{R.}~\bibnamefont{{Livi}}},\ and\ \bibinfo {author}
  {\bibfnamefont{A.}~\bibnamefont{{Politi}}},\ }%
  \bibfield{title}{%
  \enquote{\bibinfo {title} {{Characterizing Dynamics with Covariant Lyapunov
  Vectors}},}\ }%
  \bibfield{journal}{%
  \Doi{10.1103/PhysRevLett.99.130601}{\bibinfo {journal} {Physical Review
  Letters}}\ }%
  \textbf{\bibinfo {volume} {99}},\ \bibinfo {eid} {130601} (\bibinfo {month}
  {Sep.}\ \bibinfo {year} {2007}),\
  \Eprint{http://arxiv.org/abs/0706.0510}{arXiv:0706.0510 [nlin.CD]}%
  \bibAnnoteFile{NoStop}{2007PhRvL..99m0601G}%
\bibitem{2013JPhA...46y4005G}%
  \BibitemOpen
  \bibfield{author}{%
  \bibinfo {author} {\bibfnamefont{F.}~\bibnamefont{{Ginelli}}}, \bibinfo
  {author} {\bibfnamefont{H.}~\bibnamefont{{Chat{\'e}}}}, \bibinfo {author}
  {\bibfnamefont{R.}~\bibnamefont{{Livi}}},\ and\ \bibinfo {author}
  {\bibfnamefont{A.}~\bibnamefont{{Politi}}},\ }%
  \bibfield{title}{%
  \enquote{\bibinfo {title} {{Covariant Lyapunov vectors}},}\ }%
  \bibfield{journal}{%
  \Doi{10.1088/1751-8113/46/25/254005}{\bibinfo {journal} {Journal of Physics A
  Mathematical General}}\ }%
  \textbf{\bibinfo {volume} {46}},\ \bibinfo {eid} {254005} (\bibinfo {month}
  {Jun.}\ \bibinfo {year} {2013}),\
  \Eprint{http://arxiv.org/abs/1212.3961}{arXiv:1212.3961 [nlin.CD]}%
  \bibAnnoteFile{NoStop}{2013JPhA...46y4005G}%
\bibitem{MR516310}%
  \BibitemOpen
  \bibfield{author}{%
  \bibinfo {author} {\bibfnamefont{David}\ \bibnamefont{Ruelle}},\ }%
  \bibfield{title}{%
  \enquote{\bibinfo {title} {An inequality for the entropy of differentiable
  maps},}\ }%
  \bibfield{journal}{%
  \Doi{10.1007/BF02584795}{\bibinfo {journal} {Bol. Soc. Brasil. Mat.}}\ }%
  \textbf{\bibinfo {volume} {9}},\ \bibinfo {pages} {83--87} (\bibinfo {year}
  {1978}),\ ISSN \bibinfo {issn} {0100-3569}%
  \bibAnnoteFile{NoStop}{MR516310}%
\bibitem{Ruelle1989}%
  \BibitemOpen
  \bibfield{author}{%
  \bibinfo {author} {\bibfnamefont{David}\ \bibnamefont{Ruelle}},\ }%
  \emph{\bibinfo {title} {{Chaotic Evolution and Strange Attractors}}}\
  (\bibinfo {publisher} {Cambridge University Press},\ \bibinfo {address}
  {Cambridge},\ \bibinfo {year} {1989})%
  \bibAnnoteFile{NoStop}{Ruelle1989}%
\bibitem{Katok1995}%
  \BibitemOpen
  \bibfield{author}{%
  \bibinfo {author} {\bibfnamefont{Anatole}\ \bibnamefont{Katok}}\ and\
  \bibinfo {author} {\bibfnamefont{Boris}\ \bibnamefont{Hasselblatt}},\ }%
  \emph{\bibinfo {title} {{Introduction to the Modern Theory of Dynamical
  Systems}}}\ (\bibinfo {publisher} {Cambridge U. Press},\ \bibinfo {address}
  {Cambridge},\ \bibinfo {year} {1995})%
  \bibAnnoteFile{NoStop}{Katok1995}%
\bibitem{Ledrappier1985}%
  \BibitemOpen
  \bibfield{author}{%
  \bibinfo {author} {\bibfnamefont{F.}~\bibnamefont{Ledrappier}}\ and\ \bibinfo
  {author} {\bibfnamefont{L.-S.}\ \bibnamefont{Young}},\ }%
  \bibfield{title}{%
  {\selectlanguage {english}\enquote{\bibinfo {title} {The metric entropy of
  diffeomorphisms: Part ii: Relations between entropy, exponents and
  dimension},}\ }}%
  \bibfield{journal}{%
  \bibinfo {journal} {Annals of Mathematics},\ }%
  \bibinfo {series} {Second Series}\ \textbf{\bibinfo {volume} {122}},\
  \bibinfo {pages} {pp. 540--574} (\bibinfo {year} {1985}),\ ISSN \bibinfo
  {issn} {0003486X},\ \url{http://www.jstor.org/stable/1971329}%
  \bibAnnoteFile{NoStop}{Ledrappier1985}%
\bibitem{ETS:7891821}%
  \BibitemOpen
  \bibfield{author}{%
  \bibinfo {author} {\bibfnamefont{Mauro}\ \bibnamefont{Patr\~{a}o}},\ }%
  \bibfield{title}{%
  \enquote{\bibinfo {title} {Entropy and its variational principle for
  non-compact metric spaces},}\ }%
  \bibfield{journal}{%
  \Doi{10.1017/S0143385709000674}{\bibinfo {journal} {Ergodic Theory and
  Dynamical Systems}}\ }%
  \textbf{\bibinfo {volume} {30}},\ \bibinfo {pages} {1529--1542} (\bibinfo
  {month} {10}\ \bibinfo {year} {2010}),\ ISSN \bibinfo {issn} {1469-4417}%
  \bibAnnoteFile{NoStop}{ETS:7891821}%
\bibitem{1987CMaPh.112..691C}%
  \BibitemOpen
  \bibfield{author}{%
  \bibinfo {author} {\bibfnamefont{A.}~\bibnamefont{{Connes}}}, \bibinfo
  {author} {\bibfnamefont{H.}~\bibnamefont{{Narnhofer}}},\ and\ \bibinfo
  {author} {\bibfnamefont{W.}~\bibnamefont{{Thirring}}},\ }%
  \bibfield{title}{%
  \enquote{\bibinfo {title} {{Dynamical entropy of C* algebras and von Neumann
  algebras}},}\ }%
  \bibfield{journal}{%
  \Doi{10.1007/BF01225381}{\bibinfo {journal} {Communications in Mathematical
  Physics}}\ }%
  \textbf{\bibinfo {volume} {112}},\ \bibinfo {pages} {691--719} (\bibinfo
  {month} {Dec.}\ \bibinfo {year} {1987})%
  \bibAnnoteFile{NoStop}{1987CMaPh.112..691C}%
\bibitem{1994LMaPh..32...75A}%
  \BibitemOpen
  \bibfield{author}{%
  \bibinfo {author} {\bibfnamefont{R.}~\bibnamefont{{Alicki}}}\ and\ \bibinfo
  {author} {\bibfnamefont{M.}~\bibnamefont{{Fannes}}},\ }%
  \bibfield{title}{%
  \enquote{\bibinfo {title} {{Defining quantum dynamical entropy}},}\ }%
  \bibfield{journal}{%
  \Doi{10.1007/BF00761125}{\bibinfo {journal} {Letters in Mathematical
  Physics}}\ }%
  \textbf{\bibinfo {volume} {32}},\ \bibinfo {pages} {75--82} (\bibinfo {month}
  {Sep.}\ \bibinfo {year} {1994})%
  \bibAnnoteFile{NoStop}{1994LMaPh..32...75A}%
\bibitem{2006JPhA...39R..65V}%
  \BibitemOpen
  \bibfield{author}{%
  \bibinfo {author} {\bibfnamefont{A.}~\bibnamefont{{Vourdas}}},\ }%
  \bibfield{title}{%
  \enquote{\bibinfo {title} {{Topical Review: Analytic representations in
  quantum mechanics}},}\ }%
  \bibfield{journal}{%
  \Doi{10.1088/0305-4470/39/7/R01}{\bibinfo {journal} {Journal of Physics A
  Mathematical General}}\ }%
  \textbf{\bibinfo {volume} {39}},\ \bibinfo {pages} {65} (\bibinfo {month}
  {Feb.}\ \bibinfo {year} {2006})%
  \bibAnnoteFile{NoStop}{2006JPhA...39R..65V}%
\end{thebibliography}


%

\end{document}